\documentclass{article} 
\usepackage{olmoasr,times}

\usepackage{amsmath,amsfonts,bm}

\def\eqref#1{equation~\ref{#1}}

\def\1{\bm{1}}

\DeclareMathAlphabet{\mathsfit}{\encodingdefault}{\sfdefault}{m}{sl}
\SetMathAlphabet{\mathsfit}{bold}{\encodingdefault}{\sfdefault}{bx}{n}

\usepackage{hyperref}
\usepackage{url}
\usepackage{graphicx}
\usepackage{booktabs}

\newcommand{\poolname}{\textsc{OLMoASR-Pool}}
\newcommand{\datasetname}{\textsc{OLMoASR-Mix}}
\newcommand{\modelname}{\textsc{OLMoASR}}
\newcommand{\ourtiny}{\modelname-tiny.en}
\newcommand{\ourbase}{\modelname-base.en}
\newcommand{\oursmall}{\modelname-small.en}
\newcommand{\ourmedium}{\modelname-medium.en}
\newcommand{\ourlarge}{\modelname-large.en}

\newcommand{\commentout}[1]{}

\definecolor{mygreen}{RGB}{40,209,19}
\definecolor{myred}{RGB}{212,17,89}

\title{\modelname: Open Models and Data for \\ Training Robust Speech Recognition Models}

\author{Huong Ngo$^{1,2*}$ Matt Deitke Martijn Bartelds$^{3}$ Sarah Pratt$^{2}$ \\ \textbf{Josh Gardner}$^{\dagger}$ \textbf{Matt Jordan}$^{1,\dagger}$ \textbf{Ludwig Schmidt}$^{2*,3,\dagger}$ \\
$^{1}$Allen Institute for AI, $^{2}$University of Washington, $^{3}$Stanford University}

\iclrfinalcopy 
\begin{document}

\maketitle

\renewcommand{\thefootnote}{} 
\footnotetext{$^{\dagger}$Equal senior contributions. Authors are listed alphabetically by last name.}
\footnotetext{$^{*}$Part of work done while at University of Washington.}
\renewcommand{\thefootnote}{\arabic{footnote}} 

\begin{abstract}
  Improvements in training data scale and quality have led to significant advances, yet its influence in speech recognition remains underexplored. In this paper, we present a large-scale dataset, \poolname, and series of models, \modelname, to study and develop robust zero-shot speech recognition models. Beginning from \poolname, a collection of 3M hours of English audio and 17M\commentout{manually-uploaded} transcripts, we design text heuristic filters to remove low-quality or mistranscribed data. Our curation pipeline produces a new dataset containing 1M hours of high-quality audio-transcript pairs, which we call \datasetname. We use \datasetname~to train the \modelname~suite of models, ranging from 39M (tiny.en) to 1.5B (large.en) parameters. Across all model scales, \modelname{} achieves comparable average performance to OpenAI's Whisper on short and long-form speech recognition benchmarks. Notably, \ourmedium{} attains a 12.8\% and 11.0\% word error rate (WER) that is on par with Whisper's largest English-only model Whisper-medium.en's 12.4\% and 10.5\% WER for short and long-form recognition respectively (at equivalent parameter count). \poolname, \datasetname, \modelname~models, and filtering, training and evaluation code will be made publicly available to further research on robust speech processing.
\end{abstract}

\section{Introduction}
\commentout{Over the past five years,}Foundation models trained on web-scale data have changed the landscape of AI. Scaling up models for language, vision-language, and speech has led to breakthroughs such as GPT \citep{brown2020language}, CLIP \citep{radford2021learning}, and Whisper \citep{whisper}, and the\commentout{universality} generalization capabilities of these new models has enabled a wide range of new applications. Training data is key to these advances:\commentout{all three examples and subsequent} modern AI models rely on large training sets harvested from the Web that combine both broad data collection with detailed curation. For instance, the latest language models are now trained on trillions of text tokens produced by sophisticated data pipelines \citep{grattafiori2024llama3, liu2024deepseek, li2024datacomp, olmo20242, liu2023llm360}.

The importance of web-scale training data has led to increasing interest in datasets, including several efforts to build open datasets for training foundation models. In the text domain, researchers have introduced a multitude of datasets such as C4 \citep{raffel2023exploringlimitstransferlearning}, the Pile \citep{gao2020pile800gbdatasetdiverse}, RedPajama \citep{weber2024redpajamaopendatasettraining}, RefinedWeb \citep{penedo2023refinedweb}, Dolma \citep{soldaini2024dolmaopencorpustrillion}, DCLM \citep{li2025datacomplmsearchgenerationtraining}, FineWeb \citep{penedo2024fineweb}, Nemotron-CC \citep{su2025nemotroncctransformingcommoncrawl}, etc.
In addition, researchers have proposed a wide range of data curation methods \citep{li2024datacomp, su2024nemotron, penedo2024fineweb, penedo2023refinedweb, wettig2025organize}. Together, these efforts have enabled multiple open source language models that in some cases are competitive with closed-source models and serve as an important starting point for open research. Similarly, the open image-text datasets such as YFCC, LAION, and DataComp have served as a catalyst for research on multimodal learning, leading to reproductions of frontier commercial models such as OpenCLIP \citep{cherti2023reproducible}. The speech domain, however, is currently lagging behind the other modalities: where there are important efforts such as OWSM \citep{owsm-orig} and YODAS \citep{yodas}, there is currently no open-source reproduction of the full-scale Whisper \citep{whisper} models, nor a publicly available training set to begin such an effort. This is despite the widespread use and significant impact of the Whisper model\footnote{Whisper is, for example, OpenAI’s most-starred repository on GitHub, and various official versions of Whisper have garnered at least 17M downloads on Hugging Face as of the date of this publication.}, and the stated importance of large-scale, high-quality data to Whisper’s performance \citep{whisper}. 

\begin{figure}[htbp]
    \centering
    \includegraphics[scale=0.3]{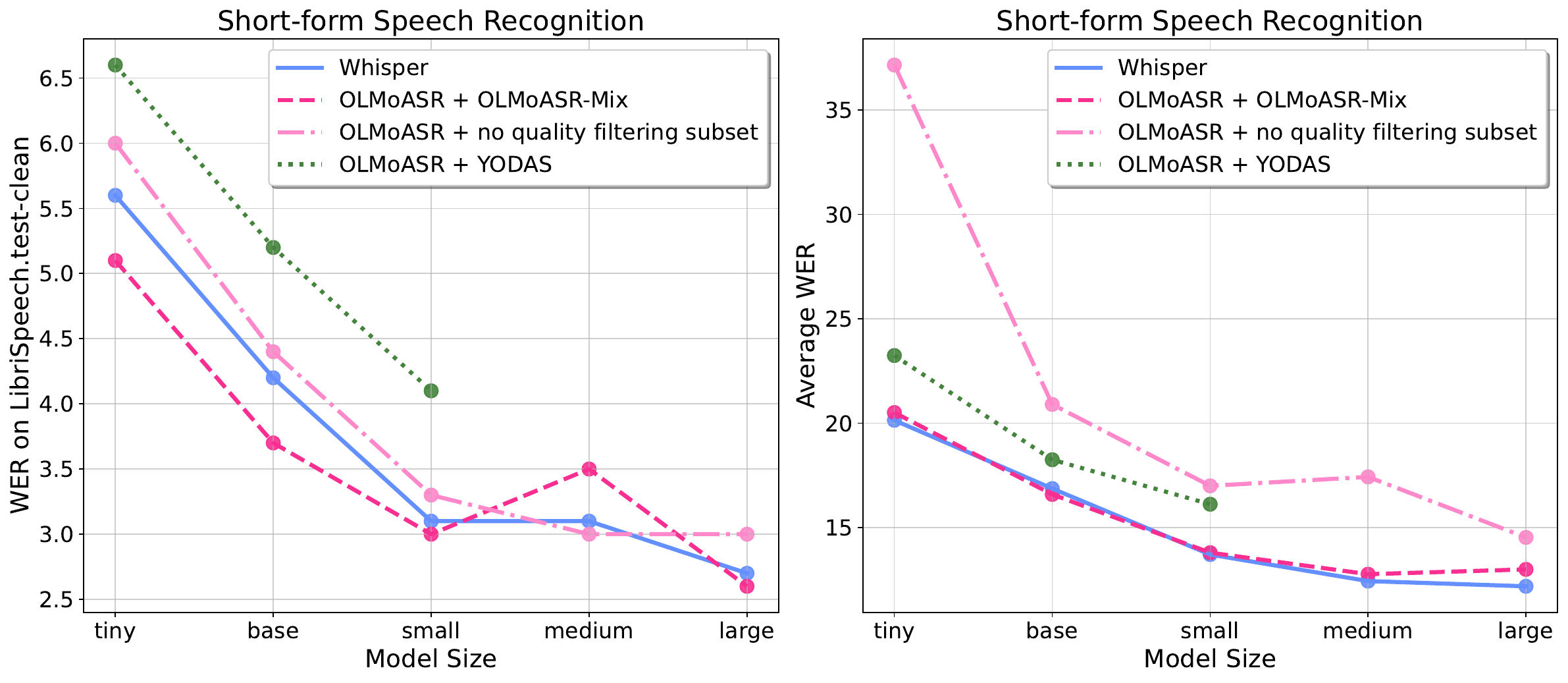}
    \caption{Performance on LibriSpeech.test-clean (left) and average performance across 14 short-form speech recognition benchmarks (right) of each baseline for all possible model scales.}
    \label{fig:short_libri_wer_combined}
\end{figure}


We address this shortcoming in the open data ecosystem by introducing \poolname{}, a dataset with 3M hours of audio and associated transcripts taken from the public internet. Starting from these audio-text pairs, we build a careful data curation pipeline that allows us to assemble a high-quality subset for training state-of-the-art robust, zero-shot speech recognition models.
As a result of this pipeline, we propose \datasetname{}, a dataset with 1M hours of audio and accompanying transcripts, that surpasses the scale of data used to train the initial Whisper models.
Our data scale matches the amount of weakly labeled data used to train the second and third versions of the Whisper models. 

We validate the quality of \datasetname{} by training a range of models following the Whisper architecture and training recipe. The resulting family of models, \modelname{}, closely matches the quality of Whisper on a wide range of benchmarks and across multiple compute scales up to the largest Whisper model scale (see Figure~\ref{fig:short_libri_wer_combined} and Figure~\ref{fig:short_long_wer}). In addition, our models outperform other open data speech recognition models such as OWSM (see Table~\ref{tab:owsm_olmoasr_short}), wav2vec, and HuBERT (see Table~\ref{tab:short_form_more}) across a range of short- and long-form speech recognition benchmarks. Our experiments show that our data curation pipeline is key to the success of our models: compared to a baseline that was trained on a filtered \poolname{} to remove non-English audio-transcript pairs, our actual training set \datasetname{} consistently improves performance across compute scales (see Figure~\ref{fig:short_long_wer}). A key step in our pipeline is removing repeating lines, which improves performance by 14.5\% WER (percentage points). A dataset quality ablation also demonstrates that training an \modelname{} model on a weakly-supervised, web-scale data collection like \datasetname{} results in better performance across many evaluation sets compared to training on academic datasets.

We publicly release the IDs of the audio-text pairs in \poolname{} and \datasetname{} as a starting point for research on speech training data. Our hope is that this will enable new research on data curation and speech recognition, similar to the LAION-5B project for multimodal learning and DataComp-LM for language modeling. In addition, providing a web-scale speech recognition training set increases transparency around current approaches to AI training and enables research on bias in datasets, fairness, privacy, and data auditing. Given the sensitive nature of training data, we strongly recommend that \poolname{} and \datasetname{} should only be used for academic research purposes in its current form. We advise against any applications in deployed systems without carefully investigating the legal, privacy, and fairness risks associated with \poolname{} and \datasetname{}.

Our training data is available at \href{https://huggingface.co/datasets/allenai/OLMoASR-Pool}{https://huggingface.co/datasets/allenai/OLMoASR-Pool}, code can be found at \href{https://github.com/allenai/OLMoASR}{https://github.com/allenai/OLMoASR} and models are available at \href{https://huggingface.co/datasets/allenai/OLMoASR}{https://huggingface.co/datasets/allenai/OLMoASR}.

\section{Data}

\subsection{Motivation}
Whisper \citep{whisper} demonstrated an approach of scaling speech recognition datasets to achieve strong generalization and robustness. While the related work focused on studying the impact of data scaling on zero-shot generalization, not much is known about the impact of dataset design and the dataset itself was never made publicly accessible.

Open Whisper-style Speech Model (OWSM) \citep{owsm-orig,  owsmctc, owsm-3.1} is an effort to reproduce Whisper with open-source tools, but trained on a mix of academic datasets. \citep{owsm-3.2} investigates the effects of data quality on OWSM models using the same mix. However, studying and training on such data pool does not enable rigorous investigation into Whisper's zero-shot capability. Moreover, the performance of those models demonstrate that despite improvements to the architecture or training recipe, dataset composition plays a central role in supplying the model's generalization and robust capabilities. 

To address this knowledge gap, we conduct experiments on a collection of weakly-supervised audio-transcript data that is on the same scale as Whisper's dataset to analyze how different dataset design choices affect a speech recognition model's downstream performance. Model architecture, training code and evaluation setup are controlled and only the data is changed. More specifically, we use the same architecture, tokenizer and evaluation setup as Whisper. As Whisper did not publish their training and data processing code, we construct a training loop and data processing pipeline to the best of our abilities to match what Whisper used. To validate that our training loop was correct, we monitor the model's training and validation loss curves.

\subsection{Curation}
\begin{figure}[htbp]
    \centering
    \includegraphics[scale=0.28]{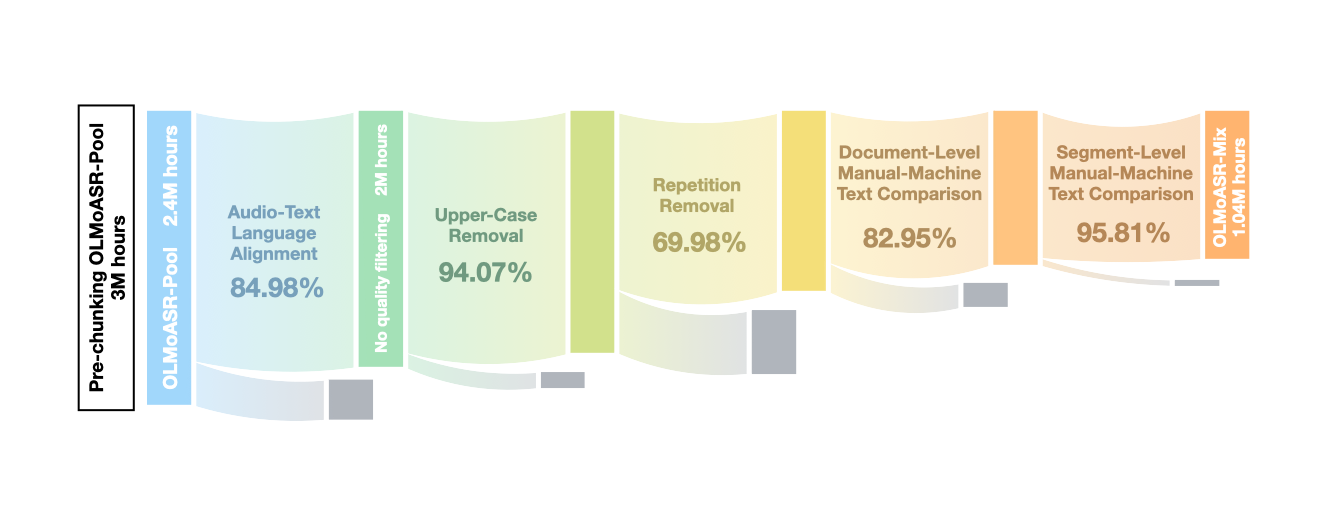}
    \caption{Construction of \datasetname~from \poolname. Segmentation reduces \poolname~from 3M to 2.4M hours. Percentages are relative to most recent filtered subset, based on the number of hours or segments.}
    \label{fig:filter_sankey}
\end{figure}

In this section, we describe the curation choices made to achieve \datasetname{} and quantify the impact of the curation layer. Firstly, to ensure that the audio and text language matches, we perform audio-text language alignment (Section~\ref{sec:alignment}). Next, we experiment with different text-based heuristics (Section~\ref{sec:heuristics}) to remove low-quality audio-text pairs. We will also explain what type of low-quality data we are targeting at each layer. Finally, we perform fuzzy decontamination and deduplication (Appendix \ref{app:dedupclean}) on the transcripts to remove contaminated or duplicated audio-text pairs. Figure~\ref{fig:filter_sankey} visually illustrates each step in removing low quality data and denotes their respective percentages of data removed.

All experiments are performed on the \ourtiny{} model and compared to a baseline that has only been trained on data filtered with the audio-text language alignment filter. We use this baseline as it does not target the quality of transcripts. This will be referred to as the "no quality filtering" baseline from this point onward. To assess them, we use the word error rate (WER) metric which calculates the percentage of words that were incorrectly predicted when compared to a reference text.

\subsubsection{Audio-text language alignment} \label{sec:alignment}
To ensure that we are training on only English audio-text pairs, a spoken language identification model, VoxLingua107~\citep{voxlingua}, is used to tag the audio sample with the spoken language, and \verb+pycld2+ to tag the corresponding transcript sample with the text language. The top-1 predicted language from both models are chosen as the tagged languages. We then remove audio-text pairs where the tagged audio and text language are not both English.

\subsubsection{Text heuristics} \label{sec:heuristics}
There is no guarantee that the transcripts are manual transcriptions of the audio from the public internet. In fact, many publicly accessible transcripts are produced by speech recognition systems. Recent work has illustrated inferior performance from training on automatic transcripts for speech recognition~\citep{yodas} or a mix of human and machine-labeled data on translation systems~\citep{fernandes2023scalinglawsmultilingualneural}. Through manual examination, we identify text characteristics of machine-generated transcripts that can be used as filtering heuristics.

Exploratory analysis uncovered a non-trivial number of audio-text pairs where the transcriptions are unfaithful and unrelated to the audio. There are also instances of partial transcriptions, and temporally misaligned transcriptions. To remove them, each audio-text pair is scored based on the WER between the manually uploaded and an associated machine-generated text, then omitting pairs where the score is lower than a specific threshold.

\paragraph{Text casing.} 
Through examining audio-text pairs, we noticed that a lot of machine-generated transcripts contain text that is mostly made up of lower or upper case characters. We designed a case detector to loop through each transcript line and keep counts of the respective cases. The case type with the highest frequency is the resulting case tag of the audio-text pair. In Table~\ref{tab:text_casing}, removing audio-text pairs that have been tagged with upper or lower case types improves short-form WER by 4.8\% after removing 32.0\% of the data.

\begin{table}[h]
\centering
\renewcommand{\arraystretch}{1} 
\makebox[\textwidth]{
\resizebox{0.7\textwidth}{!}{
\scriptsize
\begin{tabular}{lcccc}
\toprule
Filtering strategy & \parbox{1.2cm}{\centering Data hours} & \parbox{1cm}{\centering Percent remaining (\%)} & \parbox{1.2cm}{\centering Short-form WER} \\
\midrule
No quality filtering & 2,010,447 & - & 37.2 \\
Upper-case or lower-case removal & 1,367,506 & 68.0 & 32.4 \\
\bottomrule
\end{tabular}
}}
\caption{Filtering out transcripts with upper or lower case improves WER performance on short-form transcription. Short-form WER refers to the average performance across 14 short-form speech recognition datasets. Percent remaining is based on number of hours or segments remaining from filter relative to the no quality filtering strategy.}
\label{tab:text_casing}
\end{table}

\paragraph{Presence of repeating lines.}


Another issue with machine-generated transcripts is the repetition of lines, which can misalign audio and text. To detect these, we check if each line matches the previous one exactly. Table~\ref{tab:repeating_lines} shows that the removal of repeats reduces the short-form WER by 14.4\% while removing 39.9\% of the data.

We also test combining casing and repeat filters. Table~\ref{tab:repeating_lines} shows that filtering by both repeats, and lower and upper case yields a WER 0.7\% higher than just repeats and mostly uppercase text, while removing more data. Therefore, our final curation only filters based on the presence of repeating lines and mostly upper case text.

\begin{table}[h]
\centering
\renewcommand{\arraystretch}{1} 
\makebox[\textwidth]{
\resizebox{0.8\textwidth}{!}{
\scriptsize
\begin{tabular}{lcccc}
\toprule
Filtering strategy & \parbox{1.2cm}{\centering Data hours} & \parbox{1cm}{\centering Percent remaining (\%)} & \parbox{1.2cm}{\centering Short-form WER} \\
\midrule
No quality filtering & 2,010,447 & - & 37.2 \\
Repeating lines removal & 1,207,676 & 60.1 & 22.7 \\
Repeating lines removal and upper-case removal & 1,139,722 & 56.7 & 21.9 \\
Repeating lines removal and upper and lower case removal & 944,106 & 47.0 & 22.6 \\
\bottomrule
\end{tabular}
}}
\newline
\newline
\caption{Filtering out transcripts with repeating lines improves WER performance on short-form transcription. Short-form WER refers to the average performance across 14 short-form speech recognition datasets. Percent remaining is based on number of hours or segments remaining from filter relative to the no quality filtering strategy.}
\label{tab:repeating_lines}
\end{table}

\paragraph{Manual-machine text comparison.}
Unfaithful or misaligned transcripts can cause the model to learn from poorly matched audio-text pairs. To filter these, we compare a manual transcript with its machine-generated version using WER. Although automatic transcripts are less precise, they reliably capture speech utterances, making them effective for identifying low-quality data. Pairs with WER above a set threshold are removed.

We use two variants: manual-machine \textit{document-level} and \textit{segment-level} comparison. The document-level filter mainly detects unrelated transcripts, but might confound minor differences with misalignments. Manual inspection also showed that sections of poorly-aligned transcripts can be recovered, so we also utilize a segment-level filter for more fine-grained filtering. Through experiments, we determined thresholds of 0.5 for document-level and 0.7 for segment-level filtering.



Table~\ref{tab:man_mach} illustrates that employing this filter improves WER performance on short-form transcription by 16.5\% after removing 54.8\% of the data.

\begin{table}[h]
\centering
\renewcommand{\arraystretch}{1} 
\makebox[\textwidth]{
\resizebox{0.7\textwidth}{!}{
\scriptsize
\begin{tabular}{lcccc}
\toprule
Filtering strategy & \parbox{1.2cm}{\centering Data hours} & \parbox{1cm}{\centering Percent remaining (\%)} & \parbox{1.2cm}{\centering Short-form WER} \\
\midrule
No quality filtering & 2,010,447 & - & 37.2 \\
Manual-machine text comparison & 908,923 & 45.2 & 20.7 \\
\bottomrule
\end{tabular}
}}
\newline
\newline
\caption{Employing manual-machine text comparison filter improves WER performance on short-form transcription. Short-form WER refers to the average performance across 14 short-form speech recognition datasets. Percent remaining is based on number of hours or segments remaining from filter relative to the no quality filtering strategy.}
\label{tab:man_mach}
\end{table}

\section{Model and Training}

\subsection{Model}
To fully understand the impact of our data curation methodology on producing robust speech recognition systems with strong zero-shot capabilities, we utilize Whisper's model architecture and tokenizer. We have only modified the architecture code to use FlashAttention \citep{dao2022flashattentionfastmemoryefficientexact} in the attention module and incorporate the causal and padding mask for batch training. 

\subsection{Training Details}
In contrast to the Whisper training procedure, we train with a larger batch size, reconfigure the learning rate and warmup scheduler, and total steps trained accordingly. This was done to leverage available compute and maximize efficient distributed training. Moreover, we retain the same maximum learning rate for all scales and do not perform hyperparameter tuning. For \ourtiny, \ourbase{} and \oursmall{} we train with Distributed Data Parallel (DDP) using FP16 with dynamic loss scaling. In contrast, \ourmedium{} and \ourlarge{} were trained with Fully Sharded Data Parallel (FSDP) using bfloat16 with dynamic loss scaling and activation checkpointing. We found that training with FSDP using bfloat16 provided better training stability than with FP16. 

\section{Results and Discussion}

\subsection{Evaluation Methodology}
To properly assess how our dataset design approach contributes to \modelname's zero-shot generalization ability, we evaluate our model on a suite of 21 datasets that have not been used for training, 14 short-form and 7 long-form sets. To maintain comparable evaluation with Whisper \citep{whisper}, we use greedy decoding for short-form and beam search for long-form. The evaluation sets will assess the model's capabilities in contexts such as audio book recordings, lectures, calls and meetings. Moreover, these datasets contain speech of short and long utterances, different accents, high and low signal clarity. We also study how useful \datasetname~is as a robustness intervention, utilizing effective and relative robustness from \citep{taori2020measuringrobustnessnaturaldistribution}.

\begin{figure}[htbp]
    \centering
    \includegraphics[scale=0.3]{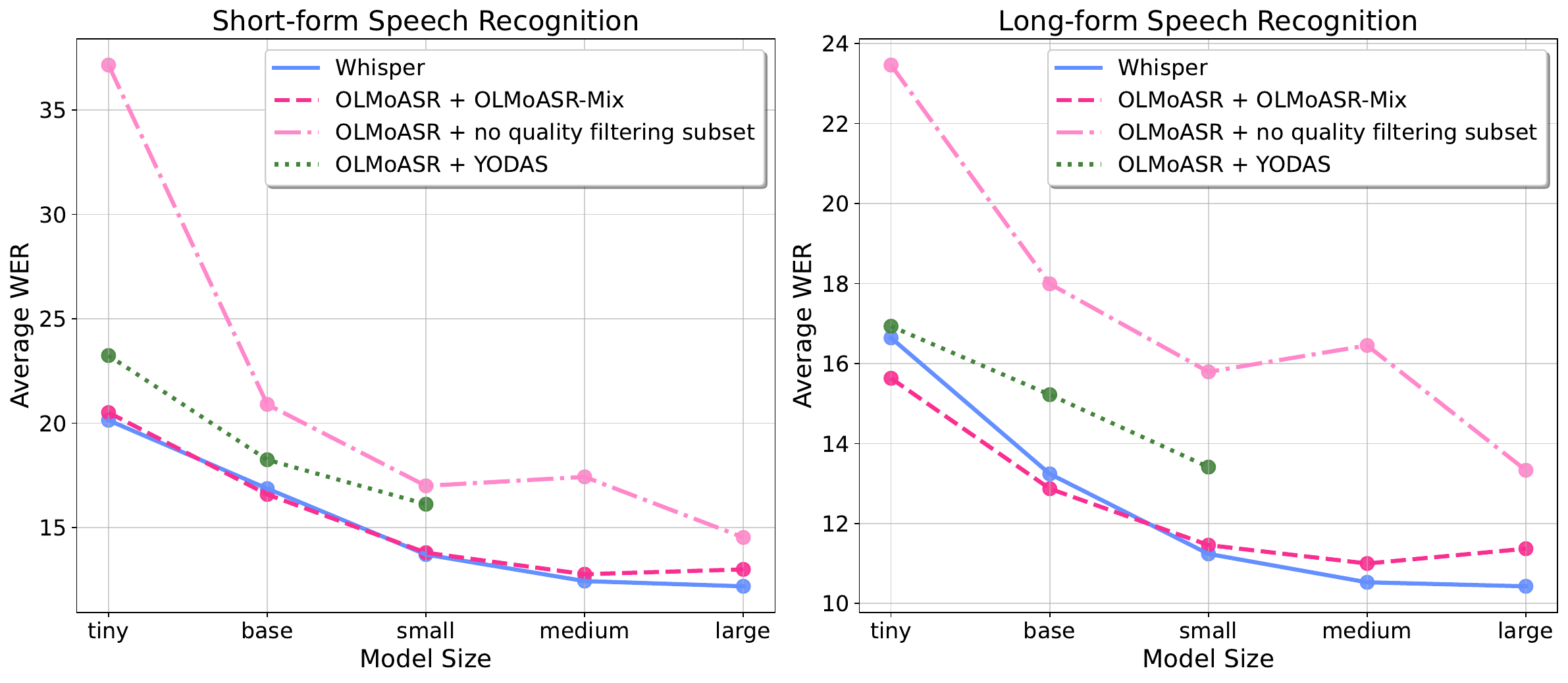}
    \caption{Average performance across 14 short-form speech recognition benchmarks (left) and across 7 long-form speech recognition benchmarks (right) of each baseline for all possible model scales.}
    \label{fig:short_long_wer}
\end{figure}

\subsection{Zero-shot performance across datasets}
\begin{table}[h]
\centering
\resizebox{0.9\textwidth}{!}{
\begin{tabular}{l*{15}{c}}
\toprule
\parbox{0.1cm}{Model} &
\rotatebox{90}{\makebox[3.1cm][c]{LibriSpeech.test-clean}} &
\rotatebox{90}{\makebox[3.1cm][c]{LibriSpeech.test-other}} &
\rotatebox{90}{\makebox[1.8cm][c]{TED-LIUM3}} &
\rotatebox{90}{\makebox[0.6cm][c]{WSJ}} &
\rotatebox{90}{\makebox[1.4cm][c]{CallHome}} &
\rotatebox{90}{\makebox[1.7cm][c]{Switchboard}} &
\rotatebox{90}{\makebox[2.45cm][c]{CommonVoice5.1}} &
\rotatebox{90}{\makebox[0.65cm][c]{Artie}} &
\rotatebox{90}{\makebox[1.35cm][c]{CORAAL}} &
\rotatebox{90}{\makebox[1.2cm][c]{CHiME6}} &
\rotatebox{90}{\makebox[1.4cm][c]{AMI-IHM}} &
\rotatebox{90}{\makebox[1.5cm][c]{AMI-SDM}} &
\rotatebox{90}{\makebox[1.8cm][c]{VoxPopuli.en}} &
\rotatebox{90}{\makebox[1.6cm][c]{Fleurs.en.us}} &
\rotatebox{90}{\makebox[1.1cm][c]{\textbf{Average}}} \\
\midrule
\multicolumn{16}{c}{\textbf{\modelname~(Open weights, code, data) vs. Whisper (Open weights, closed training code, data)}} \\ \midrule
\modelname-tiny.en & \textcolor{mygreen}{5.1} & \textcolor{mygreen}{12.3} & \textcolor{mygreen}{5.5} & 5.6 & \textcolor{mygreen}{23.9} & 18.7 & \textcolor{mygreen}{25.1} & \textcolor{mygreen}{19.3} & 25.7 & 45.2 & 24.2 & 55.4 & \textcolor{mygreen}{11.6} & \textcolor{mygreen}{9.7} & \textbf{20.5} \\
Whisper tiny.en & 5.6 & 14.6 & 6.0 & \textcolor{mygreen}{5.0} & 24.1 & \textcolor{mygreen}{17.8} & 26.3 & 20.0 & \textcolor{mygreen}{23.9} & \textcolor{mygreen}{41.3} & \textcolor{mygreen}{23.7} & \textcolor{mygreen}{50.3} & 11.7 & 11.6 & \textbf{20.1} \\
\midrule
\modelname-base.en & \textcolor{mygreen}{3.7} & \textcolor{mygreen}{9.0} & \textcolor{mygreen}{4.6} & \textcolor{mygreen}{4.3} & \textcolor{mygreen}{20.5} & \textcolor{mygreen}{14.0} & \textcolor{mygreen}{18.5} & 13.6 & \textcolor{mygreen}{21.5} & 38.0 & \textcolor{mygreen}{20.4} & 47.8 & \textcolor{mygreen}{9.7} & \textcolor{mygreen}{6.7} & \textbf{16.6} \\
Whisper base.en & 4.2 & 10.2 & 4.9 & 4.6 & 20.9 & 15.2 & 19.0 & \textcolor{mygreen}{13.4} & 22.6 & \textcolor{mygreen}{36.4} & 20.5 & \textcolor{mygreen}{46.7} & 10.0 & 7.6 & \textbf{16.9} \\
\midrule
\modelname-small.en & \textcolor{mygreen}{3.0} & \textcolor{mygreen}{7.0} & 4.2 & 3.8 & \textcolor{mygreen}{16.7} & \textcolor{mygreen}{13.2} & 13.1 & \textcolor{mygreen}{9.6} & \textcolor{mygreen}{19.6} & 30.6 & 18.7 & 39.9 & 8.7 & \textcolor{mygreen}{5.0} & \textbf{13.8} \\
Whisper small.en & 3.1 & 7.4 & \textcolor{mygreen}{4.0} & \textcolor{mygreen}{3.3} & 18.2 & 15.7 & 13.1 & 9.7 & 20.2 & \textcolor{mygreen}{27.6} & \textcolor{mygreen}{17.5} & \textcolor{mygreen}{38.0} & \textcolor{mygreen}{8.1} & \textcolor{mygreen}{6.0} & \textbf{13.7} \\
\midrule
\modelname-medium.en & 3.5 & \textcolor{mygreen}{5.7} & 5.0 & 3.6 & \textcolor{mygreen}{14.3} & \textcolor{mygreen}{12.7} & 11.3 & \textcolor{mygreen}{7.5} & 18.7 & 28.5 & 16.9 & 38.3 & 8.4 & \textcolor{mygreen}{4.4} & \textbf{12.8} \\
Whisper medium.en & \textcolor{mygreen}{3.1} & 6.3 & \textcolor{mygreen}{4.1} & \textcolor{mygreen}{3.3} & 16.2 & 14.1 & \textcolor{mygreen}{10.6} & 7.6 & \textcolor{mygreen}{17.5} & \textcolor{mygreen}{25.3} & \textcolor{mygreen}{16.4} & \textcolor{mygreen}{37.2} & \textcolor{mygreen}{7.4} & 5.0 & \textbf{12.4} \\
\midrule
\modelname-large.en & \textcolor{mygreen}{2.6} & 5.9 & 4.5 & 3.7 & 16.5 & 12.7 & 11.1 & 7.9 & 18.7 & 30.7 & 16.4 & 38.8 & 8.1 & 4.5 & \textbf{13.0} \\
\modelname-large.en-v2 & 2.7 & 5.6 & 4.2 & 3.6 & \textcolor{mygreen}{15.0} & \textcolor{mygreen}{11.7} & 11.1 & 7.8 & \textcolor{mygreen}{18.1} & 29.4 & 17.1 & 38.0 & 8.0 & \textcolor{mygreen}{4.2} & \textbf{12.6} \\
Whisper large-v1 & 2.7 & 5.6 & \textcolor{mygreen}{4.0} & \textcolor{mygreen}{3.1} & 15.8 & 13.1 & \textcolor{mygreen}{9.5} & \textcolor{mygreen}{6.7} & 19.4 & \textcolor{mygreen}{25.6} & 16.4 & \textcolor{mygreen}{36.9} & \textcolor{mygreen}{7.3} & 4.6 & \textbf{12.2}\\
\midrule
Whisper large-v2 & 2.7 & 5.2 & 4.0 & 3.9 & 17.6 & 13.8 & 9.0 & 6.2 & 16.2 & 25.5 & 16.9 & 36.4 & 7.3 & 4.4 & \textbf{12.1} \\
Whisper large-v3 & 2.0 & 3.9 & 3.9 & 3.5 & 14.0 & 13.2 & 8.4 & 5.9 & 18.7 & 26.8 & 16.0 & 34.2 & 9.5 & 4.0 & \textbf{11.7} \\
Whisper large-v3-turbo & 2.2 & 4.2 & 3.5 & 3.5 & 13.2 & 12.9 & 9.7 & 6.3 & 18.6 & 27.3 & 16.1 & 35.2 & 12.2 & 4.4 & \textbf{12.1} \\
\bottomrule
\end{tabular}}
\newline
\newline
\caption{Short-form English transcription WER (\%) with greedy decoding, comparing between \modelname~and Whisper models.}
\label{tab:short_form}
\end{table}

\paragraph{Primary results.} Average performance across 14 short and 7 long-form evaluation sets can be found in Figure~\ref{fig:short_long_wer}. Short-form performance from each dataset can be found in Table~\ref{tab:short_form} and long-form results can be found in Table~\ref{tab:long_form}. Below, we establish core findings from our main baseline.

\paragraph{\datasetname~enables \modelname's competitive zero-shot capability.} From Figure~\ref{fig:short_long_wer}, \modelname~is comparable to Whisper \citep{whisper}, the current state-of-the-art zero-shot ASR, with the largest average performance gap being 0.4\% for short-form and 1\% for long-form.

For short-form transcription, \modelname~performs on-par with Whisper at tiny to small scales. However, the gap widens at 769M and 1.5B, which may be due to lack of hyperparameter tuning or differences in data scale. Specifically, \ourlarge{} was trained on 440K hours of English data per pass, while Whisper used 680K hours of multilingual data. For a more fair comparison, we re-trained \ourlarge{} on 680K hours of English data which reduces the gap from 0.8\% to 0.4\%. This is denoted as \ourlarge{}-v2 on~\ref{tab:short_form}.

For long-form transcription, \ourtiny{} and \ourbase{} outperform Whisper’s equivalents, and \oursmall{} is on par with Whisper-small.en. At larger scales, the performance gap reappears for similar reasons as in short-form transcription.



\begin{figure}[htbp]
    \centering
    \includegraphics[scale=0.3]{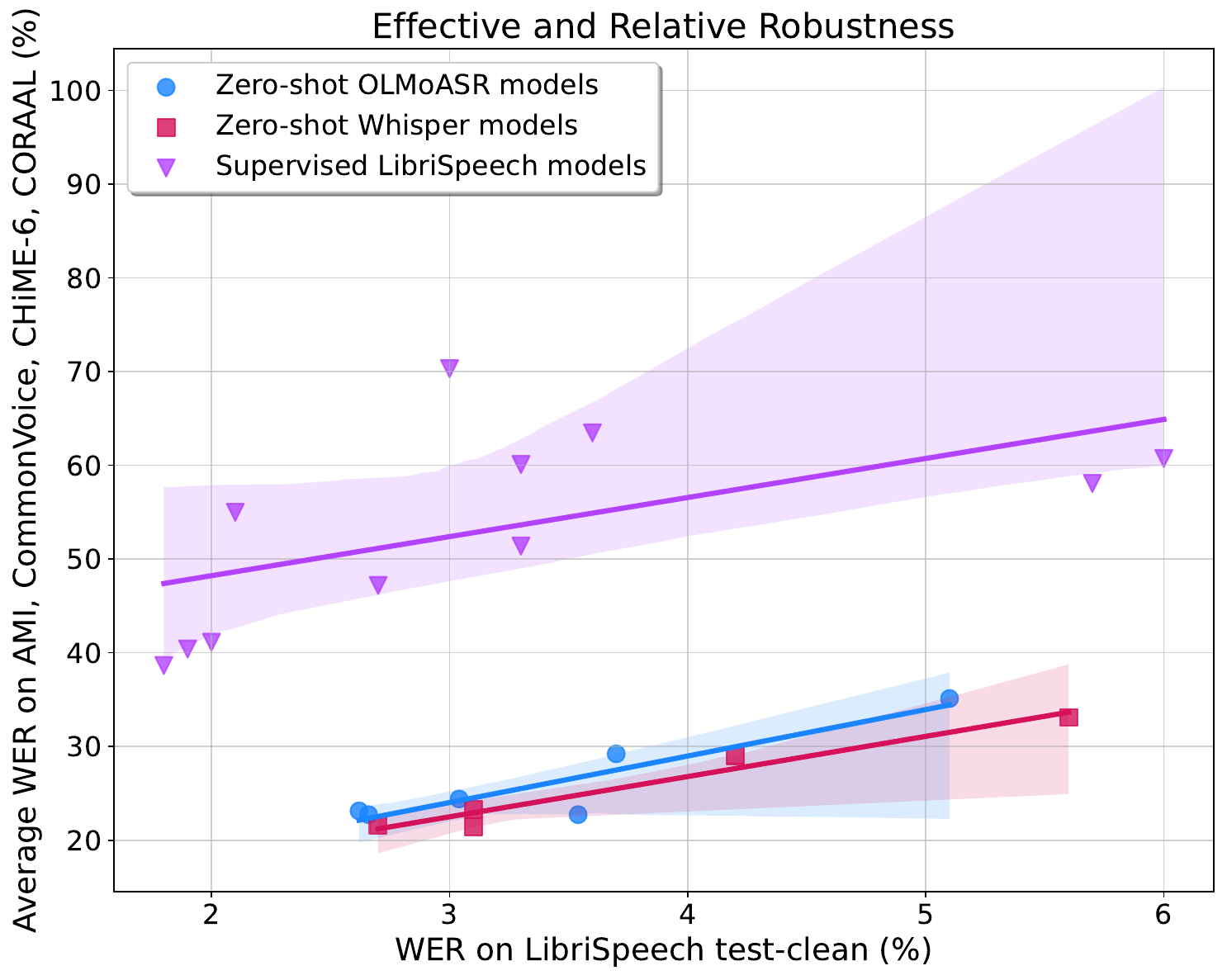}
    \caption{We plot 11 supervised models trained on LibriSpeech without any robustness interventions and demonstrate their WER on a reference test set and the average WER across 5 out-of-distribution evaluation sets. We also plot zero-shot \modelname~models to compare to the standard models, and Whisper models to demonstrate \modelname's similar robustness capability. }
    \label{fig:robustness}
\end{figure}


\paragraph{Data curation is vital to achieve strong zero-shot generalization.} \modelname~on all model scales benefits from data curation, especially \ourtiny{} for short-form and long-form, and \ourmedium{} for long-form. This can be observed from the performance discrepancy between \modelname{} trained on the no quality filtering subset and \datasetname{} on short and long-form in Figure~\ref{fig:short_long_wer}.

\begin{table}[h]
\centering
\renewcommand{\arraystretch}{1} 
\makebox[\textwidth]{
\resizebox{0.7\textwidth}{!}{
\scriptsize
\begin{tabular}{l*{8}{c}}
\toprule
\parbox{0.1cm}{Model} &
\rotatebox{90}{{TED-LIUM3}} &
\rotatebox{90}{{Meanwhile}} &
\rotatebox{90}{{Kincaid46}} &
\rotatebox{90}{{Rev16}} &
\rotatebox{90}{{Earnings-21}} &
\rotatebox{90}{{Earnings-22}} &
\rotatebox{90}{{CORAAL}} &
\rotatebox{90}{{\textbf{Average}}} \\
\midrule
\multicolumn{9}{c}{\textbf{\modelname~(Open weights, code, data) vs. Whisper (Open weights, closed training code, data)}} \\ \midrule
\modelname-tiny.en & \textcolor{mygreen}{4.8} & \textcolor{mygreen}{12.6} & \textcolor{mygreen}{13.6} & \textcolor{mygreen}{14.0} & \textcolor{mygreen}{14.2} & \textcolor{mygreen}{20.0} & \textcolor{mygreen}{30.2} & \textbf{15.6} \\
Whisper tiny.en & 5.5 & 12.8 & 13.8 & 15.1 & 17.0 & 22.0 & 30.3 & \textbf{16.6} \\
\midrule
\modelname-base.en & \textcolor{mygreen}{3.9} & 10.2 & 11.2 & \textcolor{mygreen}{12.0} & \textcolor{mygreen}{11.1} & \textcolor{mygreen}{15.6} & 26.1 & \textbf{12.9} \\
Whisper base.en & 4.6 & \textcolor{mygreen}{9.4} & 11.2 & 13.2 & 12.5 & 16.6 & \textcolor{mygreen}{25.2} & \textbf{13.2} \\
\midrule
\modelname-small.en & \textcolor{mygreen}{3.6} & 7.4 & 10.2 & \textcolor{mygreen}{11.5} & \textcolor{mygreen}{10.1} & 14.0 & 23.4 & \textbf{11.5} \\
Whisper small.en & 4.6 & \textcolor{mygreen}{6.0} & \textcolor{mygreen}{9.4} & 12.0 & 10.8 & 14.0 & \textcolor{mygreen}{21.9} & \textbf{11.2} \\
\midrule
\modelname-medium.en & \textcolor{mygreen}{3.3} & 6.9 & 9.4 & 12.5 & \textcolor{mygreen}{9.5} & 13.5 & 21.9 & \textbf{11.0} \\
Whisper medium.en & 3.6 & \textcolor{mygreen}{5.2} & \textcolor{mygreen}{8.9} & \textcolor{mygreen}{11.9} & 10.2 & \textcolor{mygreen}{13.3} & \textcolor{mygreen}{20.6} & \textbf{10.5} \\
\midrule
\modelname-large.en & \textcolor{mygreen}{3.5} & 8.8 & 10.0 & 11.5 & 9.9  & 13.5 & 22.4 & \textbf{11.4} \\
\modelname-large.en-v2 & 3.6 & 10.0 & 10.1 & 11.1 & \textcolor{mygreen}{9.8}  & 13.5 & 22.1 & \textbf{11.5} \\
Whisper large-v1 & 3.8 & \textcolor{mygreen}{5.3} & \textcolor{mygreen}{8.8} & \textcolor{mygreen}{11.0} & 10.3 & \textcolor{mygreen}{13.4} & \textcolor{mygreen}{20.4} & \textbf{10.4} \\
\midrule
Whisper large-v2 & 3.5 & 5.1 & 8.8 & 11.3 & 9.7 & 12.6 & 19.6 & \textbf{10.1} \\
Whisper large-v3 & 3.2 & 5.2 & 8.3 & 10.2 & 9.4 & 12.8 & 19.4 & \textbf{9.8} \\
Whisper large-v3-turbo & 3.1 & 5.2 & 8.4 & 9.5 & 9.5 & 12.6 & 19.3 & \textbf{9.7} \\
\bottomrule
\end{tabular}
}}
\newline
\newline
\caption{Long-form English transcription WER (\%) with beam search and temperature fallback, comparing between \modelname~and Whisper models.}
\label{tab:long_form}
\end{table}

\subsection{Robustness gained from web-scale data}

\paragraph{Effective robustness.}
Following \citep{taori2020measuringrobustnessnaturaldistribution}, effective robustness measures how much a model outperforms the expected baseline on out-of-distribution data, given its in-distribution performance. A positive gap indicates stronger robustness than a standard model.

To evaluate \modelname's effective robustness, we use LibriSpeech test-clean as the in-distribution set and five out-of-distribution sets: AMI (AMI-IHM, AMI-SDM), CommonVoice, CHiME-6, and CORAAL, covering diverse speakers and conditions.

Figure~\ref{fig:robustness} shows that although zero-shot \modelname~models have higher WER on LibriSpeech test-clean than supervised models, they significantly outperform them on the out-of-distribution benchmarks.

\paragraph{Relative robustness.} Effective robustness on its own is insufficient to characterize the robustness of a model. Hence, we use relative robustness to directly measure the performance difference between a model with and without a robustness intervention. From Figure~\ref{fig:robustness}, we can examine the relative robustness \modelname~has compared to supervised LibriSpeech models which illustrates that \modelname~out-performs the other models on the out-of-distribution datasets. 

\paragraph{\datasetname~is a useful robustness intervention.} From analyzing \modelname's effective and relative robustness, \modelname~exhibits positive effective and relative robustness, making \datasetname~and the curation methodology to extract it a beneficial robustness solution. 


\section{Ablations}

\subsection{Dataset Scaling}
\begin{figure}[htbp]
    \centering
    \includegraphics[scale=0.3]{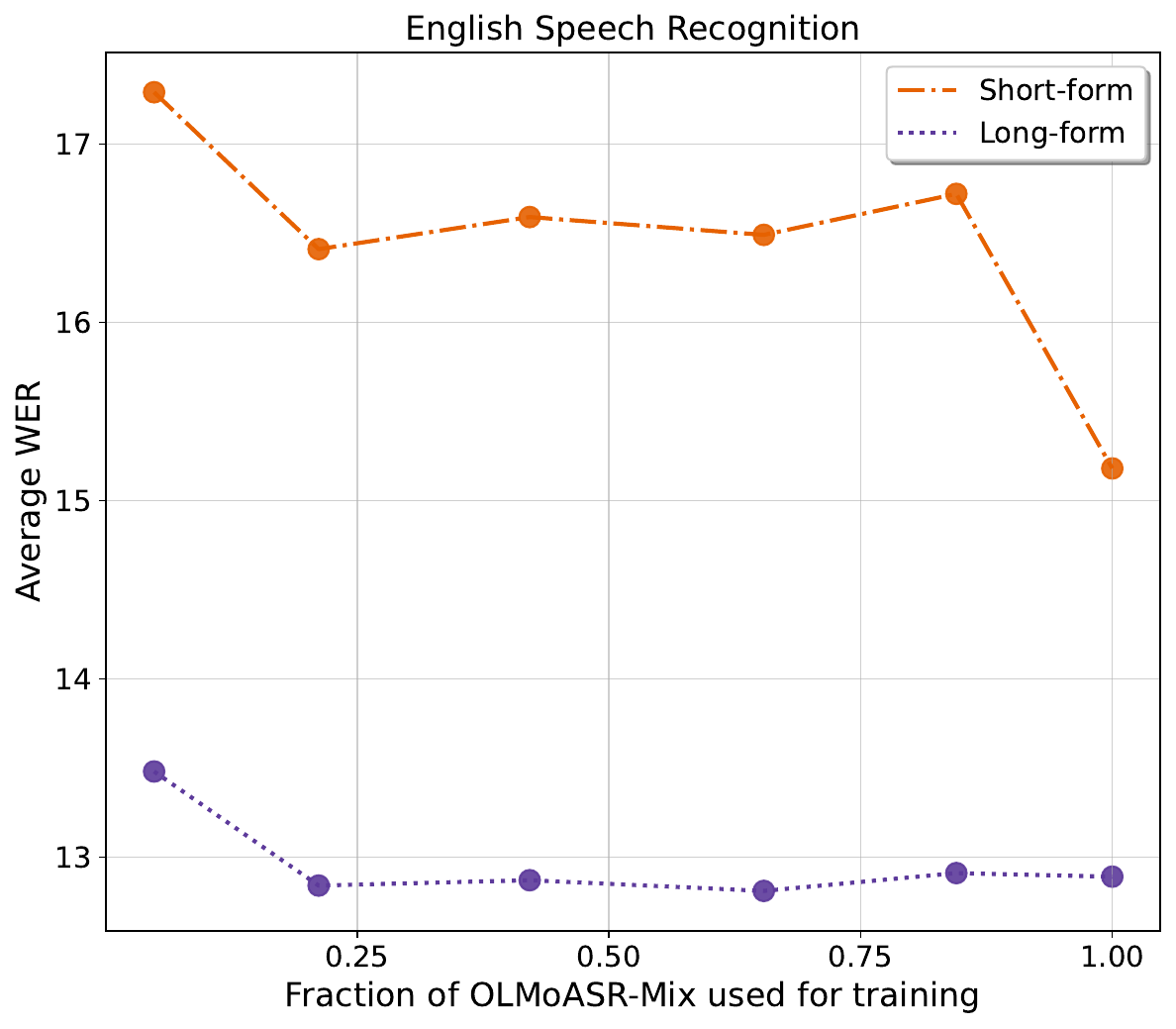}
    \caption{We plot the average performance of \modelname-74M on 14 short-form and 7 long-form evaluation sets, while varying the total data trained on. The fraction of \datasetname~used for training is based on number of hours.}
    \label{fig:data_scale}
\end{figure}

For our main experiments, we trained on 440K hours, but \datasetname~contains 1M hours. To examine the effect of data scaling, we trained \modelname-74M on subsampled portions of \datasetname: 4.8\%, 21.1\%, 42.1\%, 65.4\%, 84.5\% and 100\% (about 50K, 220K, 440K, 680K, 880K and 1M hours), keeping total seen data and hyperparameters constant.

Figure~\ref{fig:data_scale} shows that for short-form speech recognition, WER drops by 0.9\% when increasing data from 50K to 220K hours (4$\times$), but plateaus from 21.1\% to 84.5\%. Using the full dataset yields an additional 1.5\% WER improvement. For long-form, \modelname~shows minimal gains across scales. This suggests that beyond moderate scaling (1.5$\times$), gains diminish: a 20$\times$ scale-up gives only a 2.1\% WER boost for short-form and 0.6\% for long-form. This may be due to \modelname-74M being too small, the need for more data curation, longer training, or larger models.

Our work has shown that training on \datasetname~leads to strong robustness and zero-shot capabilities. Can we also quantify the performance and robustness gap between \modelname~and other models trained on a different data mix? To address this question, we evaluate \modelname~that has been trained on an academic dataset mix and \modelname~that has been trained on a dataset mix containing manual and automatic transcripts.

\subsection{Results from training on academic datasets}

\begin{figure}[htbp]
    \centering
    \includegraphics[scale=0.3]{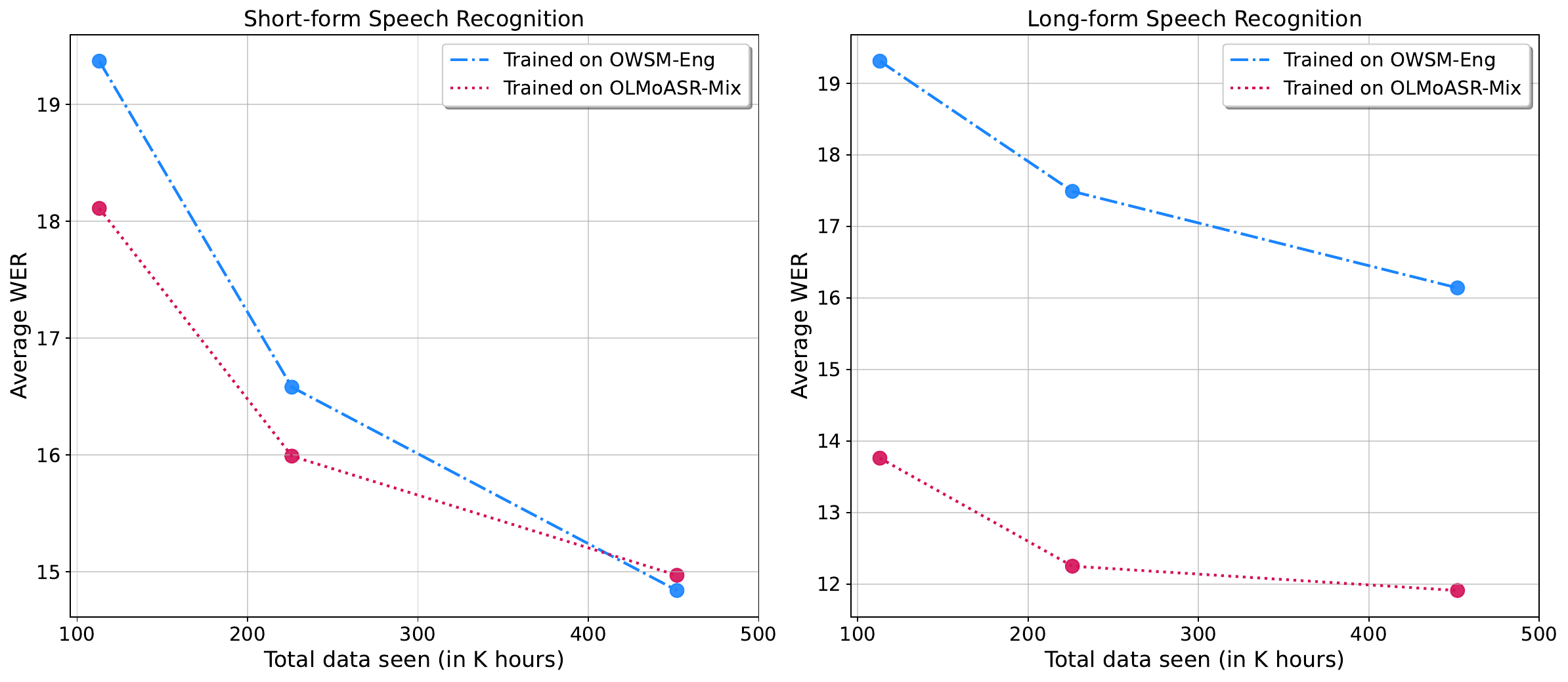}
    \caption{Short-form speech recognition performance of \modelname-244M trained on OWSM-Eng and \datasetname~for varying total amount of data seen. The baseline trained on \datasetname~trains for one epoch on subsampled subsets of the data, while \modelname-244M trained on OWSM-Eng does one, two and four passes through its 113K English subset.}
    \label{fig:owsm_olmoasr}
\end{figure}

To compare training on academic vs. web-scale data, we train \oursmall{} on OWSM-Eng (the English subset of OWSM) and \datasetname~with the same total data seen: 113K, 226K, and 452K hours and evaluate both on short and long-form speech recognition.

Figure~\ref{fig:owsm_olmoasr} shows that \datasetname~consistently yields lower WER for short-form speech except at 452K hours, where the gap is only 0.2\%. The difference is more pronounced for long-form, highlighting better generalization from unsegmented long-form data in \poolname~versus short-form academic corpora. The smaller short-form gap is partly because OWSM-Eng includes training splits of some evaluation sets, making its evaluation not fully zero-shot.

\begin{figure}[htbp]
    \centering
    \includegraphics[scale=0.28]{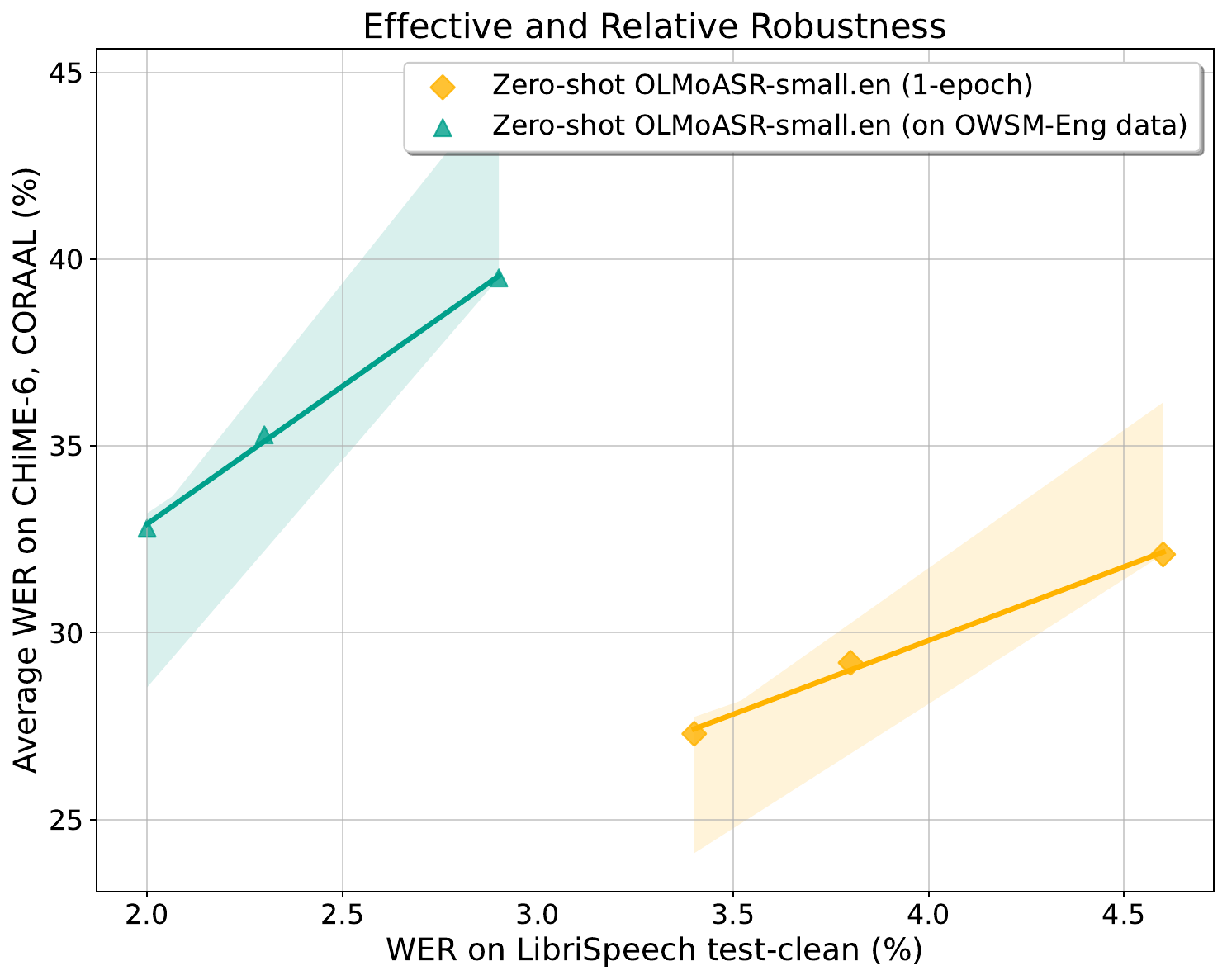}
    \caption{We plot 3 \modelname~trained on OWSM-Eng data without any robustness interventions and demonstrate their WER on a reference test set (LibriSpeech test-clean) and the average WER across 2 out-of-distribution evaluation sets (CHiME-6, CORAAL). We also plot the performance of zero-shot \modelname~models to compare to the former.}
    \label{fig:owsm_olmoasr_robustness}
\end{figure}

Since OWSM-Eng overlaps with many test sets except CHIME-6 and CORAAL, we assess out-of-distribution robustness using them and LibriSpeech as a reference. Figure~\ref{fig:owsm_olmoasr_robustness} shows that \datasetname-trained models achieve lower WER than OWSM-Eng-trained ones, outperforming expected baselines on CHIME-6 and CORAAL.

Overall, training on \datasetname~improves performance and robustness over academic data at the same scale.

\subsection{Results from training on manual and automatic data mix}
YODAS \citep{yodas} is a large-scale, multilingual speech dataset containing over 500K hours of YouTube audio across 100+ languages designed to support supervised and self-supervised learning. For our ablation, we train \modelname~with the 190K hours English subset of YODAS on model scales ranging from tiny to small. The models are trained for the same amount of total data seen as \modelname~on the full \datasetname. 

Figure~\ref{fig:short_long_wer} demonstrates that while YODAS is also a web-scale dataset, \modelname~trained on \datasetname~out-performs the YODAS-trained model on all model scales with the largest difference being 2.7\%.

\section{Related Work}

\paragraph{Large-scale English ASR Datasets}
English ASR datasets have grown dramatically in scale. LibriSpeech~\citep{librispeech} remains a benchmark with 960 hours of read speech. GigaSpeech~\citep{gigaspeech} expands this to 10,000 hours after filtering from 33,000 hours. The People's Speech~\citep{peoplesspeech} offers 30,000 hours from internet sources (excluding YouTube). Proprietary datasets like those for Whisper~\citep{whisper} and USM~\citep{googleusm} are much larger, ranging from 100K to 1M hours. YODAS~\citep{yodas} helps close this gap by providing 190,000 hours of English audio within a 480,000-hour multilingual YouTube corpus.

\paragraph{Large-scale English ASR Models}
ASR performance benefits from more and better data~\citep{w2v2, whisper}. Self-supervised learning (SSL) uses large unlabeled audio for pre-training, then fine-tunes on transcripts~\citep{bigssl, googleusm, seamless}, but fine-tuning may limit robustness~\citep{whisper}. Supervised training on diverse data enhances generalization~\citep{speechstew, likhomanenko21_interspeech}. Whisper, trained on 680K hours, is well-known but proprietary. OWSM~\citep{owsm-orig, owsm-3.1, owsm-3.2} provides an open alternative with up to 180K hours (73K English). OWLS~\citep{owls} further explores scaling laws for multilingual ASR up to 360K hours and shows clear benefits from scaling data and model size, especially for non-English.

\paragraph{Data Quality and Data-centric Learning}
Recent work across language, vision, and multimodal domains shows that better data can greatly boost model performance. Llama 2 and 3~\citep{touvron2023llama2, grattafiori2024llama3} improved mainly through better data over Llama 1~\citep{touvron2023llama1}. Similar trends hold for text (DCLM~\citep{li2024datacomp}, Nemotron-CC~\citep{su2024nemotron}) and multimodal models (DataComp~\citep{gadre2023datacomp}, DeepSeek-VL2~\citep{wu2024deepseek}, Bunny~\citep{he2024efficient}). A key principle in data-centric ML is to run \emph{controlled experiments} with fixed architectures and training, varying only the data to isolate its impact—an approach central to works like DataComp~\citep{gadre2023datacomp, li2024datacomp}.

\bibliography{olmoasr}

\begin{thebibliography}{44}
\providecommand{\natexlab}[1]{#1}
\providecommand{\url}[1]{\texttt{#1}}
\expandafter\ifx\csname urlstyle\endcsname\relax
  \providecommand{\doi}[1]{doi: #1}\else
  \providecommand{\doi}{doi: \begingroup \urlstyle{rm}\Url}\fi

\bibitem[Baevski et~al.(2020)Baevski, Zhou, Mohamed, and Auli]{w2v2}
Alexei Baevski, Yuhao Zhou, Abdelrahman Mohamed, and Michael Auli.
\newblock wav2vec 2.0: A framework for self-supervised learning of speech representations.
\newblock In H.~Larochelle, M.~Ranzato, R.~Hadsell, M.F. Balcan, and H.~Lin (eds.), \emph{Advances in Neural Information Processing Systems}, volume~33, pp.\  12449--12460. Curran Associates, Inc., 2020.
\newblock URL \url{https://proceedings.neurips.cc/paper_files/paper/2020/file/92d1e1eb1cd6f9fba3227870bb6d7f07-Paper.pdf}.

\bibitem[Brown et~al.(2020)Brown, Mann, Ryder, Subbiah, Kaplan, Dhariwal, Neelakantan, Shyam, Sastry, Askell, et~al.]{brown2020language}
Tom Brown, Benjamin Mann, Nick Ryder, Melanie Subbiah, Jared~D Kaplan, Prafulla Dhariwal, Arvind Neelakantan, Pranav Shyam, Girish Sastry, Amanda Askell, et~al.
\newblock Language models are few-shot learners.
\newblock \emph{Advances in neural information processing systems}, 33:\penalty0 1877--1901, 2020.

\bibitem[Chan et~al.(2021)Chan, Park, Lee, Zhang, Le, and Norouzi]{speechstew}
William Chan, Daniel Park, Chris Lee, Yu~Zhang, Quoc Le, and Mohammad Norouzi.
\newblock Speechstew: Simply mix all available speech recognition data to train one large neural network, 2021.
\newblock URL \url{https://arxiv.org/abs/2104.02133}.

\bibitem[Chen et~al.(2021)Chen, Chai, Wang, Du, Zhang, Weng, Su, Povey, Trmal, Zhang, Jin, Khudanpur, Watanabe, Zhao, Zou, Li, Yao, Wang, You, and Yan]{gigaspeech}
Guoguo Chen, Shuzhou Chai, Guan-Bo Wang, Jiayu Du, Wei-Qiang Zhang, Chao Weng, Dan Su, Daniel Povey, Jan Trmal, Junbo Zhang, Mingjie Jin, Sanjeev Khudanpur, Shinji Watanabe, Shuaijiang Zhao, Wei Zou, Xiangang Li, Xuchen Yao, Yongqing Wang, Zhao You, and Zhiyong Yan.
\newblock Gigaspeech: An evolving, multi-domain asr corpus with 10,000 hours of transcribed audio.
\newblock In \emph{Interspeech 2021}, pp.\  3670--3674, 2021.
\newblock \doi{10.21437/Interspeech.2021-1965}.

\bibitem[Chen et~al.(2025)Chen, Tian, Peng, Yan, Yang, and Watanabe]{owls}
William Chen, Jinchuan Tian, Yifan Peng, Brian Yan, Chao-Han~Huck Yang, and Shinji Watanabe.
\newblock Owls: Scaling laws for multilingual speech recognition and translation models, 2025.
\newblock URL \url{https://arxiv.org/abs/2502.10373}.

\bibitem[Cherti et~al.(2023)Cherti, Beaumont, Wightman, Wortsman, Ilharco, Gordon, Schuhmann, Schmidt, and Jitsev]{cherti2023reproducible}
Mehdi Cherti, Romain Beaumont, Ross Wightman, Mitchell Wortsman, Gabriel Ilharco, Cade Gordon, Christoph Schuhmann, Ludwig Schmidt, and Jenia Jitsev.
\newblock Reproducible scaling laws for contrastive language-image learning.
\newblock In \emph{Proceedings of the IEEE/CVF conference on computer vision and pattern recognition}, pp.\  2818--2829, 2023.

\bibitem[Communication et~al.(2023)Communication, Barrault, Chung, Meglioli, Dale, Dong, Duquenne, Elsahar, Gong, Heffernan, Hoffman, Klaiber, Li, Licht, Maillard, Rakotoarison, Sadagopan, Wenzek, Ye, Akula, Chen, Hachem, Ellis, Gonzalez, Haaheim, Hansanti, Howes, Huang, Hwang, Inaguma, Jain, Kalbassi, Kallet, Kulikov, Lam, Li, Ma, Mavlyutov, Peloquin, Ramadan, Ramakrishnan, Sun, Tran, Tran, Tufanov, Vogeti, Wood, Yang, Yu, Andrews, Balioglu, Costa-jussà, Celebi, Elbayad, Gao, Guzmán, Kao, Lee, Mourachko, Pino, Popuri, Ropers, Saleem, Schwenk, Tomasello, Wang, Wang, and Wang]{seamless}
Seamless Communication, Loïc Barrault, Yu-An Chung, Mariano~Cora Meglioli, David Dale, Ning Dong, Paul-Ambroise Duquenne, Hady Elsahar, Hongyu Gong, Kevin Heffernan, John Hoffman, Christopher Klaiber, Pengwei Li, Daniel Licht, Jean Maillard, Alice Rakotoarison, Kaushik~Ram Sadagopan, Guillaume Wenzek, Ethan Ye, Bapi Akula, Peng-Jen Chen, Naji~El Hachem, Brian Ellis, Gabriel~Mejia Gonzalez, Justin Haaheim, Prangthip Hansanti, Russ Howes, Bernie Huang, Min-Jae Hwang, Hirofumi Inaguma, Somya Jain, Elahe Kalbassi, Amanda Kallet, Ilia Kulikov, Janice Lam, Daniel Li, Xutai Ma, Ruslan Mavlyutov, Benjamin Peloquin, Mohamed Ramadan, Abinesh Ramakrishnan, Anna Sun, Kevin Tran, Tuan Tran, Igor Tufanov, Vish Vogeti, Carleigh Wood, Yilin Yang, Bokai Yu, Pierre Andrews, Can Balioglu, Marta~R. Costa-jussà, Onur Celebi, Maha Elbayad, Cynthia Gao, Francisco Guzmán, Justine Kao, Ann Lee, Alexandre Mourachko, Juan Pino, Sravya Popuri, Christophe Ropers, Safiyyah Saleem, Holger Schwenk, Paden Tomasello, Changhan Wang, Jeff
  Wang, and Skyler Wang.
\newblock Seamlessm4t: Massively multilingual \& multimodal machine translation, 2023.
\newblock URL \url{https://arxiv.org/abs/2308.11596}.

\bibitem[Dao et~al.(2022)Dao, Fu, Ermon, Rudra, and Ré]{dao2022flashattentionfastmemoryefficientexact}
Tri Dao, Daniel~Y. Fu, Stefano Ermon, Atri Rudra, and Christopher Ré.
\newblock Flashattention: Fast and memory-efficient exact attention with io-awareness, 2022.
\newblock URL \url{https://arxiv.org/abs/2205.14135}.

\bibitem[Fernandes et~al.(2023)Fernandes, Ghorbani, Garcia, Freitag, and Firat]{fernandes2023scalinglawsmultilingualneural}
Patrick Fernandes, Behrooz Ghorbani, Xavier Garcia, Markus Freitag, and Orhan Firat.
\newblock Scaling laws for multilingual neural machine translation, 2023.
\newblock URL \url{https://arxiv.org/abs/2302.09650}.

\bibitem[Gadre et~al.(2023)Gadre, Ilharco, Fang, Hayase, Smyrnis, Nguyen, Marten, Wortsman, Ghosh, Zhang, et~al.]{gadre2023datacomp}
Samir~Yitzhak Gadre, Gabriel Ilharco, Alex Fang, Jonathan Hayase, Georgios Smyrnis, Thao Nguyen, Ryan Marten, Mitchell Wortsman, Dhruba Ghosh, Jieyu Zhang, et~al.
\newblock Datacomp: In search of the next generation of multimodal datasets.
\newblock \emph{Advances in Neural Information Processing Systems}, 36:\penalty0 27092--27112, 2023.

\bibitem[Galvez et~al.(2021)Galvez, Diamos, Torres, Achorn, Cer\'{o}n, Gopi, Kanter, Lam, Mazumder, and Janapa~Reddi]{peoplesspeech}
Daniel Galvez, Greg Diamos, Juan Torres, Keith Achorn, Juan Cer\'{o}n, Anjali Gopi, David Kanter, Max Lam, Mark Mazumder, and Vijay Janapa~Reddi.
\newblock The people’s speech: A large-scale diverse english speech recognition dataset for commercial usage.
\newblock In J.~Vanschoren and S.~Yeung (eds.), \emph{Proceedings of the Neural Information Processing Systems Track on Datasets and Benchmarks}, volume~1, 2021.
\newblock URL \url{https://datasets-benchmarks-proceedings.neurips.cc/paper_files/paper/2021/file/202cb962ac59075b964b07152d234b70-Paper-round1.pdf}.

\bibitem[Gao et~al.(2020)Gao, Biderman, Black, Golding, Hoppe, Foster, Phang, He, Thite, Nabeshima, Presser, and Leahy]{gao2020pile800gbdatasetdiverse}
Leo Gao, Stella Biderman, Sid Black, Laurence Golding, Travis Hoppe, Charles Foster, Jason Phang, Horace He, Anish Thite, Noa Nabeshima, Shawn Presser, and Connor Leahy.
\newblock The pile: An 800gb dataset of diverse text for language modeling, 2020.
\newblock URL \url{https://arxiv.org/abs/2101.00027}.

\bibitem[Grattafiori et~al.(2024)Grattafiori, Dubey, Jauhri, Pandey, Kadian, Al-Dahle, Letman, Mathur, Schelten, Vaughan, et~al.]{grattafiori2024llama3}
Aaron Grattafiori, Abhimanyu Dubey, Abhinav Jauhri, Abhinav Pandey, Abhishek Kadian, Ahmad Al-Dahle, Aiesha Letman, Akhil Mathur, Alan Schelten, Alex Vaughan, et~al.
\newblock The llama 3 herd of models.
\newblock \emph{arXiv preprint arXiv:2407.21783}, 2024.

\bibitem[He et~al.(2024)He, Liu, Wu, Yuan, Wang, Huang, and Zhao]{he2024efficient}
Muyang He, Yexin Liu, Boya Wu, Jianhao Yuan, Yueze Wang, Tiejun Huang, and Bo~Zhao.
\newblock Efficient multimodal learning from data-centric perspective.
\newblock \emph{arXiv preprint arXiv:2402.11530}, 2024.

\bibitem[Li et~al.(2024)Li, Fang, Smyrnis, Ivgi, Jordan, Gadre, Bansal, Guha, Keh, Arora, et~al.]{li2024datacomp}
Jeffrey Li, Alex Fang, Georgios Smyrnis, Maor Ivgi, Matt Jordan, Samir~Yitzhak Gadre, Hritik Bansal, Etash Guha, Sedrick~Scott Keh, Kushal Arora, et~al.
\newblock Datacomp-lm: In search of the next generation of training sets for language models.
\newblock \emph{Advances in Neural Information Processing Systems}, 37:\penalty0 14200--14282, 2024.

\bibitem[Li et~al.(2025)Li, Fang, Smyrnis, Ivgi, Jordan, Gadre, Bansal, Guha, Keh, Arora, Garg, Xin, Muennighoff, Heckel, Mercat, Chen, Gururangan, Wortsman, Albalak, Bitton, Nezhurina, Abbas, Hsieh, Ghosh, Gardner, Kilian, Zhang, Shao, Pratt, Sanyal, Ilharco, Daras, Marathe, Gokaslan, Zhang, Chandu, Nguyen, Vasiljevic, Kakade, Song, Sanghavi, Faghri, Oh, Zettlemoyer, Lo, El-Nouby, Pouransari, Toshev, Wang, Groeneveld, Soldaini, Koh, Jitsev, Kollar, Dimakis, Carmon, Dave, Schmidt, and Shankar]{li2025datacomplmsearchgenerationtraining}
Jeffrey Li, Alex Fang, Georgios Smyrnis, Maor Ivgi, Matt Jordan, Samir Gadre, Hritik Bansal, Etash Guha, Sedrick Keh, Kushal Arora, Saurabh Garg, Rui Xin, Niklas Muennighoff, Reinhard Heckel, Jean Mercat, Mayee Chen, Suchin Gururangan, Mitchell Wortsman, Alon Albalak, Yonatan Bitton, Marianna Nezhurina, Amro Abbas, Cheng-Yu Hsieh, Dhruba Ghosh, Josh Gardner, Maciej Kilian, Hanlin Zhang, Rulin Shao, Sarah Pratt, Sunny Sanyal, Gabriel Ilharco, Giannis Daras, Kalyani Marathe, Aaron Gokaslan, Jieyu Zhang, Khyathi Chandu, Thao Nguyen, Igor Vasiljevic, Sham Kakade, Shuran Song, Sujay Sanghavi, Fartash Faghri, Sewoong Oh, Luke Zettlemoyer, Kyle Lo, Alaaeldin El-Nouby, Hadi Pouransari, Alexander Toshev, Stephanie Wang, Dirk Groeneveld, Luca Soldaini, Pang~Wei Koh, Jenia Jitsev, Thomas Kollar, Alexandros~G. Dimakis, Yair Carmon, Achal Dave, Ludwig Schmidt, and Vaishaal Shankar.
\newblock Datacomp-lm: In search of the next generation of training sets for language models, 2025.
\newblock URL \url{https://arxiv.org/abs/2406.11794}.

\bibitem[Li et~al.(2023)Li, Takamichi, Saeki, Chen, Shiota, and Watanabe]{yodas}
Xinjian Li, Shinnosuke Takamichi, Takaaki Saeki, William Chen, Sayaka Shiota, and Shinji Watanabe.
\newblock Yodas: Youtube-oriented dataset for audio and speech.
\newblock In \emph{2023 IEEE Automatic Speech Recognition and Understanding Workshop (ASRU)}, pp.\  1--8, 2023.
\newblock \doi{10.1109/ASRU57964.2023.10389689}.

\bibitem[Likhomanenko et~al.(2021)Likhomanenko, Xu, Pratap, Tomasello, Kahn, Avidov, Collobert, and Synnaeve]{likhomanenko21_interspeech}
Tatiana Likhomanenko, Qiantong Xu, Vineel Pratap, Paden Tomasello, Jacob Kahn, Gilad Avidov, Ronan Collobert, and Gabriel Synnaeve.
\newblock Rethinking evaluation in asr: Are our models robust enough?
\newblock In \emph{Interspeech 2021}, pp.\  311--315, 2021.
\newblock \doi{10.21437/Interspeech.2021-1758}.

\bibitem[Liu et~al.(2024)Liu, Feng, Xue, Wang, Wu, Lu, Zhao, Deng, Zhang, Ruan, et~al.]{liu2024deepseek}
Aixin Liu, Bei Feng, Bing Xue, Bingxuan Wang, Bochao Wu, Chengda Lu, Chenggang Zhao, Chengqi Deng, Chenyu Zhang, Chong Ruan, et~al.
\newblock Deepseek-v3 technical report.
\newblock \emph{arXiv preprint arXiv:2412.19437}, 2024.

\bibitem[Liu et~al.(2023)Liu, Qiao, Neiswanger, Wang, Tan, Tao, Li, Wang, Sun, Pangarkar, et~al.]{liu2023llm360}
Zhengzhong Liu, Aurick Qiao, Willie Neiswanger, Hongyi Wang, Bowen Tan, Tianhua Tao, Junbo Li, Yuqi Wang, Suqi Sun, Omkar Pangarkar, et~al.
\newblock Llm360: Towards fully transparent open-source llms.
\newblock \emph{arXiv preprint arXiv:2312.06550}, 2023.

\bibitem[OLMo et~al.(2024)OLMo, Walsh, Soldaini, Groeneveld, Lo, Arora, Bhagia, Gu, Huang, Jordan, et~al.]{olmo20242}
Team OLMo, Pete Walsh, Luca Soldaini, Dirk Groeneveld, Kyle Lo, Shane Arora, Akshita Bhagia, Yuling Gu, Shengyi Huang, Matt Jordan, et~al.
\newblock 2 olmo 2 furious.
\newblock \emph{arXiv preprint arXiv:2501.00656}, 2024.

\bibitem[Panayotov et~al.(2015)Panayotov, Chen, Povey, and Khudanpur]{librispeech}
Vassil Panayotov, Guoguo Chen, Daniel Povey, and Sanjeev Khudanpur.
\newblock Librispeech: An asr corpus based on public domain audio books.
\newblock In \emph{2015 IEEE International Conference on Acoustics, Speech and Signal Processing (ICASSP)}, pp.\  5206--5210, 2015.
\newblock \doi{10.1109/ICASSP.2015.7178964}.

\bibitem[Penedo et~al.(2023)Penedo, Malartic, Hesslow, Cojocaru, Cappelli, Alobeidli, Pannier, Almazrouei, and Launay]{penedo2023refinedweb}
Guilherme Penedo, Quentin Malartic, Daniel Hesslow, Ruxandra Cojocaru, Alessandro Cappelli, Hamza Alobeidli, Baptiste Pannier, Ebtesam Almazrouei, and Julien Launay.
\newblock The refinedweb dataset for falcon llm: outperforming curated corpora with web data, and web data only.
\newblock \emph{arXiv preprint arXiv:2306.01116}, 2023.

\bibitem[Penedo et~al.(2024{\natexlab{a}})Penedo, Kydl{\'\i}{\v{c}}ek, Lozhkov, Mitchell, Raffel, Von~Werra, Wolf, et~al.]{penedo2024fineweb}
Guilherme Penedo, Hynek Kydl{\'\i}{\v{c}}ek, Anton Lozhkov, Margaret Mitchell, Colin~A Raffel, Leandro Von~Werra, Thomas Wolf, et~al.
\newblock The fineweb datasets: Decanting the web for the finest text data at scale.
\newblock \emph{Advances in Neural Information Processing Systems}, 37:\penalty0 30811--30849, 2024{\natexlab{a}}.

\bibitem[Penedo et~al.(2024{\natexlab{b}})Penedo, Kydlíček, allal, Lozhkov, Mitchell, Raffel, Werra, and Wolf]{penedo2024finewebdatasetsdecantingweb}
Guilherme Penedo, Hynek Kydlíček, Loubna~Ben allal, Anton Lozhkov, Margaret Mitchell, Colin Raffel, Leandro~Von Werra, and Thomas Wolf.
\newblock The fineweb datasets: Decanting the web for the finest text data at scale, 2024{\natexlab{b}}.
\newblock URL \url{https://arxiv.org/abs/2406.17557}.

\bibitem[Peng et~al.(2023)Peng, Tian, Yan, Berrebbi, Chang, Li, Shi, Arora, Chen, Sharma, Zhang, Sudo, Shakeel, Jung, Maiti, and Watanabe]{owsm-orig}
Yifan Peng, Jinchuan Tian, Brian Yan, Dan Berrebbi, Xuankai Chang, Xinjian Li, Jiatong Shi, Siddhant Arora, William Chen, Roshan Sharma, Wangyou Zhang, Yui Sudo, Muhammad Shakeel, Jee-Weon Jung, Soumi Maiti, and Shinji Watanabe.
\newblock Reproducing whisper-style training using an open-source toolkit and publicly available data.
\newblock In \emph{2023 IEEE Automatic Speech Recognition and Understanding Workshop (ASRU)}, pp.\  1--8, 2023.
\newblock \doi{10.1109/ASRU57964.2023.10389676}.

\bibitem[Peng et~al.(2024{\natexlab{a}})Peng, Sudo, Shakeel, and Watanabe]{owsmctc}
Yifan Peng, Yui Sudo, Muhammad Shakeel, and Shinji Watanabe.
\newblock Owsm-ctc: An open encoder-only speech foundation model for speech recognition, translation, and language identification, 2024{\natexlab{a}}.
\newblock URL \url{https://arxiv.org/abs/2402.12654}.

\bibitem[Peng et~al.(2024{\natexlab{b}})Peng, Tian, Chen, Arora, Yan, Sudo, Shakeel, Choi, Shi, Chang, weon Jung, and Watanabe]{owsm-3.1}
Yifan Peng, Jinchuan Tian, William Chen, Siddhant Arora, Brian Yan, Yui Sudo, Muhammad Shakeel, Kwanghee Choi, Jiatong Shi, Xuankai Chang, Jee weon Jung, and Shinji Watanabe.
\newblock Owsm v3.1: Better and faster open whisper-style speech models based on e-branchformer.
\newblock In \emph{Interspeech 2024}, pp.\  352--356, 2024{\natexlab{b}}.
\newblock \doi{10.21437/Interspeech.2024-1194}.

\bibitem[Radford et~al.(2021)Radford, Kim, Hallacy, Ramesh, Goh, Agarwal, Sastry, Askell, Mishkin, Clark, et~al.]{radford2021learning}
Alec Radford, Jong~Wook Kim, Chris Hallacy, Aditya Ramesh, Gabriel Goh, Sandhini Agarwal, Girish Sastry, Amanda Askell, Pamela Mishkin, Jack Clark, et~al.
\newblock Learning transferable visual models from natural language supervision.
\newblock In \emph{International conference on machine learning}, pp.\  8748--8763. PmLR, 2021.

\bibitem[Radford et~al.(2023)Radford, Kim, Xu, Brockman, Mcleavey, and Sutskever]{whisper}
Alec Radford, Jong~Wook Kim, Tao Xu, Greg Brockman, Christine Mcleavey, and Ilya Sutskever.
\newblock Robust speech recognition via large-scale weak supervision.
\newblock In Andreas Krause, Emma Brunskill, Kyunghyun Cho, Barbara Engelhardt, Sivan Sabato, and Jonathan Scarlett (eds.), \emph{Proceedings of the 40th International Conference on Machine Learning}, volume 202 of \emph{Proceedings of Machine Learning Research}, pp.\  28492--28518. PMLR, 23--29 Jul 2023.
\newblock URL \url{https://proceedings.mlr.press/v202/radford23a.html}.

\bibitem[Raffel et~al.(2023)Raffel, Shazeer, Roberts, Lee, Narang, Matena, Zhou, Li, and Liu]{raffel2023exploringlimitstransferlearning}
Colin Raffel, Noam Shazeer, Adam Roberts, Katherine Lee, Sharan Narang, Michael Matena, Yanqi Zhou, Wei Li, and Peter~J. Liu.
\newblock Exploring the limits of transfer learning with a unified text-to-text transformer, 2023.
\newblock URL \url{https://arxiv.org/abs/1910.10683}.

\bibitem[Soldaini et~al.(2024)Soldaini, Kinney, Bhagia, Schwenk, Atkinson, Authur, Bogin, Chandu, Dumas, Elazar, Hofmann, Jha, Kumar, Lucy, Lyu, Lambert, Magnusson, Morrison, Muennighoff, Naik, Nam, Peters, Ravichander, Richardson, Shen, Strubell, Subramani, Tafjord, Walsh, Zettlemoyer, Smith, Hajishirzi, Beltagy, Groeneveld, Dodge, and Lo]{soldaini2024dolmaopencorpustrillion}
Luca Soldaini, Rodney Kinney, Akshita Bhagia, Dustin Schwenk, David Atkinson, Russell Authur, Ben Bogin, Khyathi Chandu, Jennifer Dumas, Yanai Elazar, Valentin Hofmann, Ananya~Harsh Jha, Sachin Kumar, Li~Lucy, Xinxi Lyu, Nathan Lambert, Ian Magnusson, Jacob Morrison, Niklas Muennighoff, Aakanksha Naik, Crystal Nam, Matthew~E. Peters, Abhilasha Ravichander, Kyle Richardson, Zejiang Shen, Emma Strubell, Nishant Subramani, Oyvind Tafjord, Pete Walsh, Luke Zettlemoyer, Noah~A. Smith, Hannaneh Hajishirzi, Iz~Beltagy, Dirk Groeneveld, Jesse Dodge, and Kyle Lo.
\newblock Dolma: an open corpus of three trillion tokens for language model pretraining research, 2024.
\newblock URL \url{https://arxiv.org/abs/2402.00159}.

\bibitem[Su et~al.(2024)Su, Kong, Lin, Jennings, Norick, Kliegl, Patwary, Shoeybi, and Catanzaro]{su2024nemotron}
Dan Su, Kezhi Kong, Ying Lin, Joseph Jennings, Brandon Norick, Markus Kliegl, Mostofa Patwary, Mohammad Shoeybi, and Bryan Catanzaro.
\newblock Nemotron-cc: Transforming common crawl into a refined long-horizon pretraining dataset.
\newblock \emph{arXiv preprint arXiv:2412.02595}, 2024.

\bibitem[Su et~al.(2025)Su, Kong, Lin, Jennings, Norick, Kliegl, Patwary, Shoeybi, and Catanzaro]{su2025nemotroncctransformingcommoncrawl}
Dan Su, Kezhi Kong, Ying Lin, Joseph Jennings, Brandon Norick, Markus Kliegl, Mostofa Patwary, Mohammad Shoeybi, and Bryan Catanzaro.
\newblock Nemotron-cc: Transforming common crawl into a refined long-horizon pretraining dataset, 2025.
\newblock URL \url{https://arxiv.org/abs/2412.02595}.

\bibitem[Taori et~al.(2020)Taori, Dave, Shankar, Carlini, Recht, and Schmidt]{taori2020measuringrobustnessnaturaldistribution}
Rohan Taori, Achal Dave, Vaishaal Shankar, Nicholas Carlini, Benjamin Recht, and Ludwig Schmidt.
\newblock Measuring robustness to natural distribution shifts in image classification, 2020.
\newblock URL \url{https://arxiv.org/abs/2007.00644}.

\bibitem[Tian et~al.(2024)Tian, Peng, Chen, Choi, Livescu, and Watanabe]{owsm-3.2}
Jinchuan Tian, Yifan Peng, William Chen, Kwanghee Choi, Karen Livescu, and Shinji Watanabe.
\newblock On the effects of heterogeneous data sources on speech-to-text foundation models.
\newblock In \emph{Interspeech 2024}, pp.\  3959--3963, 2024.
\newblock \doi{10.21437/Interspeech.2024-1938}.

\bibitem[Touvron et~al.(2023{\natexlab{a}})Touvron, Lavril, Izacard, Martinet, Lachaux, Lacroix, Rozi{\`e}re, Goyal, Hambro, Azhar, et~al.]{touvron2023llama1}
Hugo Touvron, Thibaut Lavril, Gautier Izacard, Xavier Martinet, Marie-Anne Lachaux, Timoth{\'e}e Lacroix, Baptiste Rozi{\`e}re, Naman Goyal, Eric Hambro, Faisal Azhar, et~al.
\newblock Llama: Open and efficient foundation language models.
\newblock \emph{arXiv preprint arXiv:2302.13971}, 2023{\natexlab{a}}.

\bibitem[Touvron et~al.(2023{\natexlab{b}})Touvron, Martin, Stone, Albert, Almahairi, Babaei, Bashlykov, Batra, Bhargava, Bhosale, et~al.]{touvron2023llama2}
Hugo Touvron, Louis Martin, Kevin Stone, Peter Albert, Amjad Almahairi, Yasmine Babaei, Nikolay Bashlykov, Soumya Batra, Prajjwal Bhargava, Shruti Bhosale, et~al.
\newblock Llama 2: Open foundation and fine-tuned chat models.
\newblock \emph{arXiv preprint arXiv:2307.09288}, 2023{\natexlab{b}}.

\bibitem[Valk \& Alumäe(2021)Valk and Alumäe]{voxlingua}
Jörgen Valk and Tanel Alumäe.
\newblock Voxlingua107: A dataset for spoken language recognition.
\newblock In \emph{2021 IEEE Spoken Language Technology Workshop (SLT)}, pp.\  652--658, 2021.
\newblock \doi{10.1109/SLT48900.2021.9383459}.

\bibitem[Weber et~al.(2024)Weber, Fu, Anthony, Oren, Adams, Alexandrov, Lyu, Nguyen, Yao, Adams, Athiwaratkun, Chalamala, Chen, Ryabinin, Dao, Liang, Ré, Rish, and Zhang]{weber2024redpajamaopendatasettraining}
Maurice Weber, Daniel Fu, Quentin Anthony, Yonatan Oren, Shane Adams, Anton Alexandrov, Xiaozhong Lyu, Huu Nguyen, Xiaozhe Yao, Virginia Adams, Ben Athiwaratkun, Rahul Chalamala, Kezhen Chen, Max Ryabinin, Tri Dao, Percy Liang, Christopher Ré, Irina Rish, and Ce~Zhang.
\newblock Redpajama: an open dataset for training large language models, 2024.
\newblock URL \url{https://arxiv.org/abs/2411.12372}.

\bibitem[Wettig et~al.(2025)Wettig, Lo, Min, Hajishirzi, Chen, and Soldaini]{wettig2025organize}
Alexander Wettig, Kyle Lo, Sewon Min, Hannaneh Hajishirzi, Danqi Chen, and Luca Soldaini.
\newblock Organize the web: Constructing domains enhances pre-training data curation.
\newblock \emph{arXiv preprint arXiv:2502.10341}, 2025.

\bibitem[Wu et~al.(2024)Wu, Chen, Pan, Liu, Liu, Dai, Gao, Ma, Wu, Wang, et~al.]{wu2024deepseek}
Zhiyu Wu, Xiaokang Chen, Zizheng Pan, Xingchao Liu, Wen Liu, Damai Dai, Huazuo Gao, Yiyang Ma, Chengyue Wu, Bingxuan Wang, et~al.
\newblock Deepseek-vl2: Mixture-of-experts vision-language models for advanced multimodal understanding.
\newblock \emph{arXiv preprint arXiv:2412.10302}, 2024.

\bibitem[Zhang et~al.(2022)Zhang, Park, Han, Qin, Gulati, Shor, Jansen, Xu, Huang, Wang, Zhou, Li, Ma, Chan, Yu, Wang, Cao, Sim, Ramabhadran, Sainath, Beaufays, Chen, Le, Chiu, Pang, and Wu]{bigssl}
Yu~Zhang, Daniel~S. Park, Wei Han, James Qin, Anmol Gulati, Joel Shor, Aren Jansen, Yuanzhong Xu, Yanping Huang, Shibo Wang, Zongwei Zhou, Bo~Li, Min Ma, William Chan, Jiahui Yu, Yongqiang Wang, Liangliang Cao, Khe~Chai Sim, Bhuvana Ramabhadran, Tara~N. Sainath, Françoise Beaufays, Zhifeng Chen, Quoc~V. Le, Chung-Cheng Chiu, Ruoming Pang, and Yonghui Wu.
\newblock Bigssl: Exploring the frontier of large-scale semi-supervised learning for automatic speech recognition.
\newblock \emph{IEEE Journal of Selected Topics in Signal Processing}, 16\penalty0 (6):\penalty0 1519--1532, 2022.
\newblock \doi{10.1109/JSTSP.2022.3182537}.

\bibitem[Zhang et~al.(2023)Zhang, Han, Qin, Wang, Bapna, Chen, Chen, Li, Axelrod, Wang, Meng, Hu, Rosenberg, Prabhavalkar, Park, Haghani, Riesa, Perng, Soltau, Strohman, Ramabhadran, Sainath, Moreno, Chiu, Schalkwyk, Beaufays, and Wu]{googleusm}
Yu~Zhang, Wei Han, James Qin, Yongqiang Wang, Ankur Bapna, Zhehuai Chen, Nanxin Chen, Bo~Li, Vera Axelrod, Gary Wang, Zhong Meng, Ke~Hu, Andrew Rosenberg, Rohit Prabhavalkar, Daniel~S. Park, Parisa Haghani, Jason Riesa, Ginger Perng, Hagen Soltau, Trevor Strohman, Bhuvana Ramabhadran, Tara Sainath, Pedro Moreno, Chung-Cheng Chiu, Johan Schalkwyk, Françoise Beaufays, and Yonghui Wu.
\newblock Google usm: Scaling automatic speech recognition beyond 100 languages, 2023.
\newblock URL \url{https://arxiv.org/abs/2303.01037}.

\end{thebibliography}
\bibliographystyle{olmoasr}



\appendix


\section{Training Details}
Table~\ref{tab:hyperparams} displays hyperparameters used to train all models. We trained \ourtiny{}, \ourbase{} and \oursmall{} on 1 H100 node and \ourmedium{}, \ourlarge{}, and \ourlarge{}-v2 on 2 and 4 H100 nodes respectively. 
\begin{table}[h]
\centering
\begin{tabular}{lc}
\hline
Hyperparameter & Value \\
\hline
Updates & 524288 \\
Batch Size & 512 \\
Warmup Updates & 1049 \\
Max grad norm & 1.0 \\
Optimizer & AdamW \\
$\beta_1$ & 0.9 \\
$\beta_2$ & 0.98 \\
$\epsilon$ & $10^{-6}$ \\
Weight Decay & 0.1 \\
Weight Init & Gaussian Fan-In \\
Maximum Learning Rate & $1.5 \times 10^{-3}$ \\
Learning Rate Schedule & Linear Decay \\
\hline
\end{tabular}
\caption{Training hyperparameters.}
\label{tab:hyperparams}
\end{table}

\section{Model Sizes}
Table~\ref{tab:model-sizes} enumerates all the model sizes \modelname{} has and the associated parameter count.
\begin{table}[h]
\centering
\begin{tabular}{lc}
\hline
Size & Parameters \\
\hline
tiny      & 39 M   \\
base      & 74 M   \\
small     & 244 M  \\
medium    & 769 M  \\
large     & 1550 M \\
large-v2  & 1550 M \\
\hline
\end{tabular}
\caption{Model sizes and their parameter counts}
\label{tab:model-sizes}
\end{table}

\section{Deduplication and Decontamination}
\label{app:dedupclean}
We performed transcript level fuzzy deduplication using minhash. We used the parameters from FineWeb \citep{penedo2024finewebdatasetsdecantingweb}, where we used 5-grams of tokens and computed 112 hash functions, split into 14 buckets of 8 hashes each. If any pair of transcripts has the same 8 hashes in any one bucket, they are marked as duplicates. This procedure targets documents that have a Jaccard similarity of 75\%. We performed this on 17M total transcripts and removed 505K transcripts for a total deduplication removal rate of 3\%.
Decontamination was performed by a simple n-gram search. In particular, we decontaminate the evaluation datasets of TED-LIUM3 against our training corpus. First we collect all n-grams of size 10 from the evaluation dataset and check for their presence in each training dataset transcript. If any n-gram is present, we mark the training document as contaminated and do not include it in our training sets. We apply this procedure to 17M transcripts and find only 286 contaminated transcripts.

\section{OWSM vs. OLMoASR Performance Table}
Table~\ref{tab:owsm_olmoasr_short} shows the WER performance on short-form evaluation sets that \modelname, Whisper and OWSM have all been evaluated on.
\begin{table}[h]
\centering
\makebox[\textwidth][c]{
\resizebox{0.8\textwidth}{!}{
\begin{tabular}{l*{8}{c}}
\toprule
\parbox{0.1cm}{Model} &
\rotatebox{90}{\makebox[3.1cm][c]{LibriSpeech.test-clean}} &
\rotatebox{90}{\makebox[3.1cm][c]{LibriSpeech.test-other}} &
\rotatebox{90}{\makebox[1.8cm][c]{TED-LIUM3}} &
\rotatebox{90}{\makebox[0.6cm][c]{WSJ}} &
\rotatebox{90}{\makebox[1.7cm][c]{Switchboard}} &
\rotatebox{90}{\makebox[1.8cm][c]{VoxPopuli.en}} &
\rotatebox{90}{\makebox[1.6cm][c]{Fleurs.en.us}} &
\rotatebox{90}{\makebox[1.1cm][c]{\textbf{Average}}} \\
\midrule
\ourbase & 3.7 & \textcolor{mygreen}{9.0} & \textcolor{mygreen}{4.6} & \textcolor{mygreen}{4.3} & \textcolor{mygreen}{14.0} & \textcolor{mygreen}{9.7} & \textcolor{mygreen}{6.7} & \textbf{7.4} \\
Whisper base.en & 4.2 & 10.2 & 4.9 & 4.6 & 15.2 & 10.0 & 7.6 & \textbf{8.1} \\
OWSM-v3.1 base & \textcolor{mygreen}{3.6} & 9.1 & 7.8 & 5.3 & 22.9 & 12.0 & 14.8 & \textbf{10.1} \\
\midrule
\oursmall & 3.0 & 7.0 & 4.2 & 3.8 & \textcolor{mygreen}{13.2} & 8.7 & \textcolor{mygreen}{5.0} & \textbf{6.4} \\
Whisper small.en & 3.1 & 7.4 & \textcolor{mygreen}{4.0} & \textcolor{mygreen}{3.3} & 15.7 & 8.1 & 6.0 & \textbf{6.8} \\
OWSM-v3.1 small & \textcolor{mygreen}{2.5} & \textcolor{mygreen}{5.8} & 5.0 & 3.8 & 17.4 & 9.1 & 10.3 & \textbf{7.3} \\
OWSM-v3.2 small & \textcolor{mygreen}{2.5} & 6.2 & 5.4 & 4.0 & 17.4 & 9.0 & 10.1 & \textbf{7.4} \\
\midrule
\ourmedium & 3.5 & 5.7 & 5.0 & 3.6 & \textcolor{mygreen}{12.7} & 8.4 & \textcolor{mygreen}{4.4} & \textbf{6.2} \\
Whisper medium.en & 3.1 & 6.3 & \textcolor{mygreen}{4.1} & \textcolor{mygreen}{3.3} & 14.1 & \textcolor{mygreen}{7.4} & 5.0 & \textbf{6.2} \\
OWSM-v3.1 medium & \textcolor{mygreen}{2.4} & \textcolor{mygreen}{5.0} & 5.1 & 3.5 & 16.3 & 8.4 & 9.0 & \textbf{6.8} \\
OWSM-CTC medium & \textcolor{mygreen}{2.4} & 5.2 & 4.9 & 4.2 & 16.9 & 8.6 & 9.9 & \textbf{7.0} \\
\bottomrule
\end{tabular}}}
\newline
\newline
\caption{Short-form English transcription WER (\%) with greedy decoding, comparing between \modelname, Whisper and OWSM models.}
\label{tab:owsm_olmoasr_short}
\end{table}

\section{OLMoASR vs. Other Open-source Models}
Table~\ref{tab:short_form_more} illustrates the WER performance of \modelname~relative to other open-source models.
\begin{table}[h]
\centering
\resizebox{\textwidth}{!}{
\begin{tabular}{l*{15}{c}}
\toprule
\parbox{0.1cm}{Model} &
\rotatebox{90}{\makebox[3.1cm][c]{LibriSpeech.test-clean}} &
\rotatebox{90}{\makebox[3.1cm][c]{LibriSpeech.test-other}} &
\rotatebox{90}{\makebox[1.8cm][c]{TED-LIUM3}} &
\rotatebox{90}{\makebox[0.6cm][c]{WSJ}} &
\rotatebox{90}{\makebox[1.4cm][c]{CallHome}} &
\rotatebox{90}{\makebox[1.7cm][c]{Switchboard}} &
\rotatebox{90}{\makebox[2.45cm][c]{CommonVoice5.1}} &
\rotatebox{90}{\makebox[0.65cm][c]{Artie}} &
\rotatebox{90}{\makebox[1.35cm][c]{CORAAL}} &
\rotatebox{90}{\makebox[1.2cm][c]{CHiME6}} &
\rotatebox{90}{\makebox[1.4cm][c]{AMI-IHM}} &
\rotatebox{90}{\makebox[1.5cm][c]{AMI-SDM}} &
\rotatebox{90}{\makebox[1.8cm][c]{VoxPopuli.en}} &
\rotatebox{90}{\makebox[1.6cm][c]{Fleurs.en.us}} &
\rotatebox{90}{\makebox[1.1cm][c]{\textbf{Average}}} \\
\midrule
\multicolumn{16}{c}{\textbf{\modelname~(Open weights, code, data) vs. Whisper (Open weights, closed training code, data)}} \\ \midrule
\modelname-tiny.en & \textcolor{mygreen}{5.1} & \textcolor{mygreen}{12.3} & \textcolor{mygreen}{5.5} & 5.6 & \textcolor{mygreen}{23.9} & 18.7 & \textcolor{mygreen}{25.1} & \textcolor{mygreen}{19.3} & 25.7 & 45.2 & 24.2 & 55.4 & \textcolor{mygreen}{11.6} & \textcolor{mygreen}{9.7} & \textbf{20.5} \\
Whisper tiny.en & 5.6 & 14.6 & 6.0 & \textcolor{mygreen}{5.0} & 24.1 & \textcolor{mygreen}{17.8} & 26.3 & 20.0 & \textcolor{mygreen}{23.9} & \textcolor{mygreen}{41.3} & \textcolor{mygreen}{23.7} & \textcolor{mygreen}{50.3} & 11.7 & 11.6 & \textbf{20.1} \\
\midrule
\modelname-base.en & \textcolor{mygreen}{3.7} & \textcolor{mygreen}{9.0} & \textcolor{mygreen}{4.6} & \textcolor{mygreen}{4.3} & \textcolor{mygreen}{20.5} & \textcolor{mygreen}{14.0} & \textcolor{mygreen}{18.5} & 13.6 & \textcolor{mygreen}{21.5} & 38.0 & \textcolor{mygreen}{20.4} & 47.8 & \textcolor{mygreen}{9.7} & \textcolor{mygreen}{6.7} & \textbf{16.6} \\
Whisper base.en & 4.2 & 10.2 & 4.9 & 4.6 & 20.9 & 15.2 & 19.0 & \textcolor{mygreen}{13.4} & 22.6 & \textcolor{mygreen}{36.4} & 20.5 & \textcolor{mygreen}{46.7} & 10.0 & 7.6 & \textbf{16.9} \\
\midrule
\modelname-small.en & \textcolor{mygreen}{3.0} & \textcolor{mygreen}{7.0} & 4.2 & 3.8 & \textcolor{mygreen}{16.7} & \textcolor{mygreen}{13.2} & 13.1 & \textcolor{mygreen}{9.6} & \textcolor{mygreen}{19.6} & 30.6 & 18.7 & 39.9 & 8.7 & \textcolor{mygreen}{5.0} & \textbf{13.8} \\
Whisper small.en & 3.1 & 7.4 & \textcolor{mygreen}{4.0} & \textcolor{mygreen}{3.3} & 18.2 & 15.7 & 13.1 & 9.7 & 20.2 & \textcolor{mygreen}{27.6} & \textcolor{mygreen}{17.5} & \textcolor{mygreen}{38.0} & \textcolor{mygreen}{8.1} & \textcolor{mygreen}{6.0} & \textbf{13.7} \\
\midrule
\modelname-medium.en & 3.5 & \textcolor{mygreen}{5.7} & 5.0 & 3.6 & \textcolor{mygreen}{14.3} & \textcolor{mygreen}{12.7} & 11.3 & \textcolor{mygreen}{7.5} & 18.7 & 28.5 & 16.9 & 38.3 & 8.4 & \textcolor{mygreen}{4.4} & \textbf{12.8} \\
Whisper medium.en & \textcolor{mygreen}{3.1} & 6.3 & \textcolor{mygreen}{4.1} & \textcolor{mygreen}{3.3} & 16.2 & 14.1 & \textcolor{mygreen}{10.6} & 7.6 & \textcolor{mygreen}{17.5} & \textcolor{mygreen}{25.3} & \textcolor{mygreen}{16.4} & \textcolor{mygreen}{37.2} & \textcolor{mygreen}{7.4} & 5.0 & \textbf{12.4} \\
\midrule
\modelname-large.en & \textcolor{mygreen}{2.6} & 5.9 & 4.5 & 3.7 & 16.5 & 12.7 & 11.1 & 7.9 & 18.7 & 30.7 & 16.4 & 38.8 & 8.1 & 4.5 & \textbf{13.0} \\
\modelname-large.en-v2 & 2.7 & 5.6 & 4.2 & 3.6 & \textcolor{mygreen}{15.0} & \textcolor{mygreen}{11.7} & 11.1 & 7.8 & \textcolor{mygreen}{18.1} & 29.4 & 17.1 & 38.0 & 8.0 & \textcolor{mygreen}{4.2} & \textbf{12.6} \\
Whisper large-v1 & 2.7 & 5.6 & \textcolor{mygreen}{4.0} & \textcolor{mygreen}{3.1} & 15.8 & 13.1 & \textcolor{mygreen}{9.5} & \textcolor{mygreen}{6.7} & 19.4 & \textcolor{mygreen}{25.6} & 16.4 & \textcolor{mygreen}{36.9} & \textcolor{mygreen}{7.3} & 4.6 & \textbf{12.2}\\
\midrule
Whisper large-v2 & 2.7 & 5.2 & 4.0 & 3.9 & 17.6 & 13.8 & 9.0 & 6.2 & 16.2 & 25.5 & 16.9 & 36.4 & 7.3 & 4.4 & \textbf{12.1} \\
Whisper large-v3 & 2.0 & 3.9 & 3.9 & 3.5 & 14.0 & 13.2 & 8.4 & 5.9 & 18.7 & 26.8 & 16.0 & 34.2 & 9.5 & 4.0 & \textbf{11.7} \\
Whisper large-v3-turbo & 2.2 & 4.2 & 3.5 & 3.5 & 13.2 & 12.9 & 9.7 & 6.3 & 18.6 & 27.3 & 16.1 & 35.2 & 12.2 & 4.4 & \textbf{12.1} \\
\midrule
wav2vec2-base-100h & 6.0 & 13.4 & 17.8 & 13.9 & 46.9 & 40.2 & 47.4 & 40.8 & 47.0 & 79.9 & 48.1 & 81.2 & 28.9 & 23.1 & \textbf{38.2} \\
wav2vec2-base-960h & 3.3 & 8.5 & 12.8 & 8.9 & 40.6 & 32.9 & 36.4 & 30.9 & 39.9 & 68.5 & 40.2 & 71.9 & 21.4 & 17.4 & \textbf{31.0} \\
wav2vec2-large-960h-lv60-self & 1.8 & 3.8 & 7.4 & 4.4 & 29.1 & 22.2 & 19.9 & 15.8 & 29.2 & 56.3 & 30.8 & 57.0 & 13.0 & 10.2 & \textbf{21.5} \\
wav2vec2-large-960h & 2.7 & 6.2 & 10.5 & 7.7 & 34.8 & 28.3 & 29.9 & 24.5 & 35.6 & 65.8 & 37.0 & 67.6 & 17.9 & 14.6 & \textbf{27.4} \\
wav2vec2-large-robust-ft-libri-960h & 2.6 & 5.3 & 9.2 & 6.1 & 23.4 & 19.8 & 20.3 & 16.2 & 29.4 & 58.1 & 31.7 & 61.6 & 15.1 & 11.8 & \textbf{22.2} \\
asr-crdnn-rnnlm-librispeech & 3.0 & 9.7 & 17.7 & 10.7 & 59.7 & 56.1 & 43.7 & 33.3 & 83.8 & 81.0 & 57.2 & 85.8 & 30.6 & 32.4 & \textbf{43.2} \\
asr-transformer-transformerlm-librispeech & 2.1 & 5.4 & 11.9 & 7.4 & 38.9 & 33.0 & 30.6 & 23.5 & 44.9 & 79.5 & 44.5 & 75.4 & 17.8 & 17.0 & \textbf{30.9} \\
hubert-large-ls960-ft & 2.0 & 4.1 & 8.4 & 5.4 & 29.6 & 22.8 & 20.8 & 16.0 & 32.0 & 60.0 & 33.7 & 59.1 & 14.4 & 10.9 & \textbf{22.8} \\
hubert-xlarge-ls960-ft & 1.9 & 3.5 & 8.3 & 5.4 & 29.3 & 22.2 & 19.8 & 14.8 & 31.5 & 58.5 & 33.3 & 58.9 & 14.2 & 10.5 & \textbf{22.3} \\
s2t-large-librispeech-asr & 3.3 & 8.1 & 14.9 & 9.4 & 54.5 & 40.3 & 38.1 & 30.7 & 50.2 & 79.2 & 53.4 & 79.5 & 21.6 & 18.0 & \textbf{35.8} \\
s2t-medium-librispeech-asr & 3.6 & 8.2 & 15.7 & 9.7 & 58.1 & 42.4 & 39.3 & 31.3 & 52.6 & 79.8 & 60.3 & 85.3 & 22.9 & 19.7 & \textbf{37.8} \\
stt\_en\_conformer\_ctc\_large & 2.1 & 4.2 & 4.4 & 2.1 & 11.3 & 8.2 & 7.4 & 4.0 & 13.5 & 30.5 & 15.9 & 39.9 & 6.7 & 8.2 & \textbf{11.3} \\
stt\_en\_conformer\_transducer\_xlarge & 1.5 & 2.8 & 4.3 & 1.2 & 12.0 & 7.4 & 4.3 & 1.5 & 19.9 & 36.8 & 20.5 & 48.6 & 6.0 & 6.3 & \textbf{12.4} \\
unispeech-sat-base-100h-libri-ft & 5.7 & 13.8 & 17.7 & 13.6 & 46.5 & 40.0 & 45.3 & 38.6 & 44.7 & 74.8 & 47.8 & 77.7 & 29.8 & 22.4 & \textbf{37.0} \\
\bottomrule
\end{tabular}}
\newline
\newline
\caption{Short-form English transcription WER (\%) with greedy decoding, comparing between \modelname~, Whisper models and other open-source models}
\label{tab:short_form_more}
\end{table}




\section{Contributions}
\begin{itemize}
    \item \textbf{Huong Ngo}: Programed and designed all main code infrastructures (data collection and processing, training, and evaluation), collected and processed all data, planned and executed all experiments on Ai2 compute cluster and UW Hyak Supercomputer Cluster, coordinated and performed evaluation and paper writing and revision.
    \item \textbf{Matt Deitke}: Provided advice on data collection and processing and training, paper writing and revision.
    \item \textbf{Martijn Bartelds}: Provided advice on training and evaluation, paper writing and revision.
    \item \textbf{Sarah Pratt}: Provided visual graphics support for paper.
    \item \textbf{Josh Gardner}: Provided advice on data collection and processing, model training, experiment design and project direction, paper writing and revision.
    \item \textbf{Matt Jordan}: Performed deduplication and decontamination, provided advice on data processing, experiment design and project direction, paper writing and revision.
    \item \textbf{Ludwig Schmidt}: Provided advice on data collection and processing, model training, experiment design and project direction, paper writing and revision.
\end{itemize}

\end{document}